\documentclass[journal,11pt,draftclsnofoot,twoside,onecolumn]{IEEEtran}
\usepackage{cite}

\ifCLASSINFOpdf
  \usepackage[pdftex]{graphicx}
  \DeclareGraphicsExtensions{.pdf,.jpeg,.png}
\else
  \usepackage[dvips]{graphicx}
  \DeclareGraphicsExtensions{.eps,.ps}
\fi
\usepackage[cmex10]{amsmath}
\usepackage{amsfonts}
\usepackage{amsthm}
\usepackage{array}

\usepackage[tight,footnotesize]{subfigure}
\usepackage{url}

\usepackage{bm}
\usepackage{verbatim}
\usepackage{color}
\newtheorem{theorem}{Theorem}
\newtheorem{prop}{Proposition}
\newtheorem{lemma}{Lemma}
\newtheorem{corollary}{Corollary}
\theoremstyle{remark}
\newtheorem*{remark}{Remark}

\providecommand{\abs}[1]{\left\lvert#1\right\rvert}

\DeclareMathOperator{\E}{\mathbb{E}}

\DeclareMathOperator{\var}{var}
\DeclareMathOperator{\mbbP}{\mathbb{P}}

\newcommand{\mbs}{\mathbf{s}}

\newcommand{\mbI}{\mathbf{I}}

\newcommand{\mbx}{\mathbf{x}}

\newcommand{\mbmu}{\boldsymbol{\mu}}

\newcommand{\mby}{\mathbf{y}}

\newcommand{\mbp}{\mathbf{p}}

\newcommand{\mbY}{\mathbf{Y}}
\newcommand{\mblambda}{\boldsymbol{\lambda}}

\newcommand{\mbsigma}{\boldsymbol{\sigma}}
\newcommand{\ones}{\mathbf{1}}

\newcommand{\xh}{\hat{x}}
\newcommand{\mbxh}{\hat{\mbx}}

\newcommand{\ulG}{\underline{G}}

\newcommand{\mcY}{\mathcal{Y}}

\newcommand{\mcN}{\mathcal{N}}

\newcommand{\eff}{\mathrm{eff}}

\newcommand{\SNR}{\mathrm{SNR}}

\newcommand{\na}{\mathrm{na}}
\newcommand{\BC}{\mathrm{BC}}

\begin{document}
%
\title{Performance Guarantees for Adaptive Estimation of Sparse Signals}
%
%
%

\author{Dennis~Wei~
        and~Alfred~O.~Hero,~III
\thanks{
This work was partially supported by Army Research Office grant W911NF-11-1-0391. 

D.~Wei is with the Thomas~J.~Watson Research Center, IBM Research, Yorktown Heights, NY 10598, USA, e-mail: dwei@us.ibm.com.  

A.~O.~Hero is with the Department of Electrical Engineering and Computer Science, University of Michigan, Ann Arbor, MI 48109, USA, e-mail: hero@eecs.umich.edu.}}
\maketitle

\begin{abstract}
This paper studies adaptive sensing for estimating the nonzero amplitudes of a sparse signal with the aim of providing analytical guarantees on the performance gain due to adaptive resource allocation.  We consider a previously proposed optimal two-stage policy for allocating sensing resources.  For positive powers $q$, we derive tight upper bounds on the mean $q$th-power error resulting from the optimal two-stage policy and corresponding lower bounds on the improvement over non-adaptive uniform sensing.  It is shown that the adaptation gain is related to the detectability of nonzero signal components as characterized by Chernoff coefficients, thus quantifying analytically the dependence on the sparsity level of the signal, the signal-to-noise ratio, and the sensing resource budget.  For fixed sparsity levels and increasing signal-to-noise ratio or sensing budget, we obtain the rate of convergence to oracle performance and the rate at which the fraction of resources spent on the first exploratory stage decreases to zero.  For a vanishing fraction of nonzero components, the gain increases without bound as a function of signal-to-noise ratio and sensing budget.  Numerical simulations demonstrate that the bounds on adaptation gain are quite tight in non-asymptotic regimes as well. 
\end{abstract}

\begin{IEEEkeywords}
Adaptive sensing, adaptive estimation, sparse signals, resource allocation, performance analysis.
\end{IEEEkeywords}

%
\IEEEpeerreviewmaketitle


\section{Introduction}
\label{sec:intro}

Adaptive or controlled sensing refers to the control of the signal acquisition process in response to information that has already been learned about the signal.  Recently, attention has been focused on adaptive sensing for sparse signals \cite{adapEstJSTSP2013,bashan2008,hitchings2010,bashan2011,haupt2011,malloy2011b,tajer2012,iwen2012,haupt2012,malloy2012,ji2008,castro2008}, i.e., signals that occupy a small number of dimensions in a much larger ambient space.  The common theme in these works is that the signal-to-noise ratio (SNR) can be improved for various inference tasks by gradually determining the support of the signal and allocating sensing resources accordingly. 

In \cite{adapEstJSTSP2013}, adaptive sensing is considered for the estimation of the nonzero amplitudes of a sparse signal.  For the case of two sensing stages, policies for allocating sensing resources are derived that are optimal for a variety of estimation loss functions, generalizing an optimal two-stage policy proposed by \cite{bashan2008} under slightly different motivations.  For more than two stages, \cite{adapEstJSTSP2013} also provides allocation policies based on the approximate dynamic programming method of open-loop feedback control (OLFC) \cite{bertsekas2005} with performance that improves monotonically as the number of stages increases.  Simplifications to the two-stage policy in \cite{bashan2008} have also been proposed based on Lagrangian constraint relaxation \cite{hitchings2010} and a pooling approach to reduce the number of measurements in the first stage \cite{bashan2011}. 

Empirical results in \cite{bashan2008,adapEstJSTSP2013,bashan2011} show that multistage adaptive sensing can lead to dramatically better estimates of nonzero signal components compared to non-adaptive sensing.  However, analytical quantification of the gains 
in this setting has so far been lacking.  In this paper, we make a contribution in this direction by obtaining upper bounds on the mean estimation error resulting from optimal two-stage adaptive sensing, and by extension lower bounds on the adaptation gain.  While our focus is on two-stage policies, 
the bounds also apply to the multistage OLFC policies of \cite{adapEstJSTSP2013} because of the monotonicity property noted above. 

In related work on adaptive sensing of sparse signals, performance guarantees have been derived for other inference tasks, notably signal support recovery.  Under a Gaussian observation model similar to the one in this work, bounds are obtained on the SNR required by a procedure known as distilled sensing \cite{haupt2011} to recover the support with vanishing false discovery and non-discovery rates as the signal dimension increases.  The method of sequential thresholding \cite{malloy2011b} generalizes distilled sensing and is shown to recover the support exactly in the high-dimensional limit provided that the number of observations grows as a function of the sparsity level and the SNR as measured by a Kullback-Leibler divergence.  Distilled sensing has also been generalized to a gamma-distributed observation model appropriate for spectrum sensing and the detection error probability is characterized in terms of the channel occupancy, primary user power, and sensing budget \cite{tajer2012}.  Additionally, similar support recovery guarantees have been given for sensing that is both adaptive and compressive \cite{iwen2012,haupt2012,malloy2012}, in contrast to the non-compressive component-wise observations considered in \cite{haupt2011,malloy2011b,tajer2012} and herein.  However, as contrasted with support recovery, guarantees for adaptive amplitude estimation of sparse signals have received less attention. 

This paper studies 
the benefits of adaptive sensing for the estimation of nonzero signal amplitudes in a context similar to \cite{adapEstJSTSP2013}.  
In contrast to \cite{adapEstJSTSP2013,bashan2008,hitchings2010,bashan2011} however, where the focus was on developing tractable resource allocation policies capable of large gains and on empirical validation thereof, here the focus is on theoretical certification.  Specifically, the goal is to provide estimation performance guarantees for adaptive sensing policies and to analyze the key factors affecting the adaptation gain.  Moreover, 
the guarantees presented herein are quantitative, as opposed to the weaker qualitative statements in \cite{adapEstJSTSP2013} of one policy improving upon another.

For the signal estimation problem, intuition suggests that the benefit 
of adaptive sensing depends on the extent to which sensing resources can be concentrated on the signal support, which depends in turn on the ability to detect nonzero signal components.  
It is of interest therefore to understand the impact on performance 
of several quantities:  the sparsity level of the signal, which limits the degree of concentration; the SNR, which controls detectability; and the total sensing resource budget.  
Our bounds show that 
these dependences 
can be largely summarized by a single measure of detectability, specifically a Chernoff coefficient between two distributions:  (1) the conditional distribution of the observations given the presence of a signal component; and (2) the 
marginal distribution of the observations.  For the case of mean squared error (MSE), 
the Chernoff coefficient reduces to the Bhattacharyya coefficient.  It is interesting to 
note that both Bhattacharyya \cite{jenkins2010} and Chernoff coefficients \cite{jenkins2011} have been used before in adaptive sensor management, specifically to bound the probability of classification error and assess the value of further observations. 

A more detailed overview of our results is as follows.
We focus on optimal two-stage 
policies and derive upper bounds on the mean $q$th-power estimation error for any positive $q$.  These bounds are then compared to the error incurred under non-adaptive uniform sensing.  
Two regimes are considered with respect to sparsity: (1) a fixed fraction $p \in [0,1]$ of nonzero components and signal dimension $N$ either finite or tending to infinity (Theorem \ref{thm:pFixed}); and (2) a vanishing nonzero fraction $p \to 0$ as $N \to \infty$ such that $Np \to \infty$ (Corollary \ref{cor:pVanishing}).  
To describe the combined effect of the SNR and sensing budget, in Section \ref{subsec:policyNonAdapOracle} we define a parameter $r$ as the product of SNR (in units of squared amplitude/power) and sensing budget.  In terms of $r$, the performance bounds are shown to be tight in the limits $r \to 0$ and $r \to \infty$ (note that these limits can be approached when only one of the SNR or sensing budget tends to $0$ or $\infty$).
In the case of fixed $p$ and $r \to \infty$, the analysis confirms the previous empirical finding \cite{adapEstJSTSP2013,bashan2008} that adaptive sensing can achieve the same performance as the oracle policy, which has perfect prior knowledge of the signal support.  In addition, we obtain in Theorem \ref{thm:pFixedHighSNR} rates of convergence to the oracle gain as a function of $r$ and the corresponding optimal allocation of resources to the first, exploratory stage of the two-stage procedure.  As $r \to \infty$ with $p$ fixed, the first-stage allocation decays to zero, a phenomenon that has previously been shown for two-stage variable selection and prediction \cite{firouzi2013} and is analogous to the notion of sublinear regret for multi-arm bandits (see e.g.~\cite{auer2002,audibert2007}).  In the case of vanishing $p$ and $r \to \infty$, Theorem \ref{thm:pVanishing} shows that the gain increases without bound as $O\left(\sqrt{r}\right)$ and gives the optimal division of resources between the two stages. 

The tightness of the bounds on adaptation gain is illustrated by simulation in Section \ref{sec:numVal}.  Notably, the simulations indicate that the bounds are also quite tight in the regime of intermediate $r$ and finite $N$.  Furthermore, 
while the main contribution of this paper is on performance analysis, as a by-product we propose 
a simplification of the optimal two-stage policy that minimizes an upper bound on the expected cost-to-go instead of the true cost-to-go.
The simplified policy performs nearly identically to the optimal policy, due in part to the accuracy of the bound, and unlike \cite{adapEstJSTSP2013,bashan2008,hitchings2010,bashan2011}, does not require Monte Carlo sampling to compute expectations.
  
The remainder of the paper proceeds as follows.  Sections \ref{sec:prob} and \ref{sec:policy} review relevant material from \cite{adapEstJSTSP2013}, including the signal and observation models, the estimation cost function, and a two-stage policy for sensing resource allocation that minimizes the mean estimation error.  In Section \ref{sec:results}, we develop the main results of this paper, namely bounds on the estimation error of the two-stage policy in Section \ref{sec:policy} and its improvement compared to non-adaptive sensing.  
The bounds are validated numerically in Section \ref{sec:numVal} and the paper concludes in Section \ref{sec:concl}.  Detailed proofs can be found in the appendices.

\section{Problem formulation}
\label{sec:prob}

The adaptive sensing problem considered in this paper is a special case of the multistage resource allocation problem in \cite{adapEstJSTSP2013}.  It is also related to \cite{bashan2008} as discussed in \cite{adapEstJSTSP2013}. Let $\mbx$ be an $N$-dimensional real-valued signal that is sparse in some fixed basis, which we assume is the canonical basis without loss of generality.  We consider the following Bernoulli-Gaussian model for $\mbx$:  The support of $\mbx$ is represented by a vector $\mbI$ of i.i.d.~Bernoulli indicator variables $I_{i}$, $i=1,\ldots,N$, with $\mbbP(I_{i}=1) = p$ a priori and $x_{i} = 0$ if $I_{i} = 0$.  The parameter $p$ is therefore the mean proportion of nonzero entries in $\mbx$, i.e., the mean number of nonzero entries divided by $N$.  Given $I_{i} = 1$, the corresponding amplitude $x_{i}$ is conditionally distributed 
as a Gaussian random variable with mean $\mu$ and variance $\sigma^{2}$, independent of other components of $\mbx$.  While the framework of \cite{adapEstJSTSP2013} allowed the prior distributions $\mbbP(I_{i}=1)$ and $f(x_{i} \vert I_{i}=1)$ to depend on $i$, i.e., inhomogeneous distributions, the results in the present work are restricted to the homogeneous distributions specified above.  
Furthermore, it is assumed that the prior parameters $p$, $\mu$, and $\sigma^{2}$ are known.  For the analysis herein, these parameters determine the sparsity and SNR levels that in turn control the gain of adaptive estimation.  In terms of implementation, considerable robustness to inaccurate knowledge of these parameters has been demonstrated through simulations in \cite{adapEstJSTSP2013}.

While \cite{adapEstJSTSP2013} considered the allocation of resources over an arbitrary number of stages $T$ to measure the signal $\mbx$, in this paper we focus on the two-stage case $T=2$.  
Define $\lambda_{i}(t)$, $i=1,\ldots,N, \; t = 0,\ldots,T-1,$ to be the effort allocated to sensing component $i$ of $\mbx$ in stage $t+1$, where effort may represent observation time, number of samples, or other sensing resources depending on the application.  The effort allocations $\mblambda(t) = (\lambda_{1}(t), \ldots, \lambda_{N}(t))$, $t = 0, \dots, T-1,$ 
satisfy an overall budget constraint.  Denoting by $\Lambda(t)$ the sensing budget available at the end of stage $t$, this budget constraint may be expressed as 
\begin{equation}\label{eqn:budget}
\sum_{t=0}^{T-1} \sum_{i=1}^{N} \lambda_{i}(t) = \Lambda(0),
\end{equation}
where $\Lambda(0)$ is the total budget at the beginning.
Given $\lambda_{i}(t-1)$, the observation $y_{i}(t)$ of component $i$ in stage $t$ is given by  
\begin{equation}\label{eqn:obs}
y_{i}(t) = x_{i} + \frac{n_{i}(t)}{\sqrt{\lambda_{i}(t-1)}}, \quad i=1,\ldots,N, \quad t = 1,\ldots,T,
\end{equation}
where $n_{i}(t)$ denotes i.i.d.~zero-mean Gaussian noise with variance $\nu^{2}$ and we adopt the convention that the observation is not taken if $\lambda_{i}(t-1) = 0$.  
As with $p$, $\mu$, and $\sigma^{2}$, the noise variance $\nu^{2}$ is assumed to be known and similar comments as above apply.
The effective SNR under \eqref{eqn:obs} is
\begin{equation}\label{eqn:SNReff}
\SNR_{\eff,i}(t) = \frac{\sigma^{2} \lambda_{i}(t-1)}{\nu^{2}}
\end{equation}
%
and can be selectively increased through choice of $\lambda_{i}(t-1)$. 
In the non-adaptive case, $\mblambda(t)$ for all $t$ must be chosen ahead of the first observation stage, whereas in adaptive sensing, $\mblambda(t)$ can depend causally on all observations $\mbY(t) \equiv \{\mby(1), \ldots, \mby(t)\}$ seen so far.
The function relating $\mbY(t)$ to $\mblambda(t)$ is referred to as the effort allocation policy.  
The observation model \eqref{eqn:obs}, \eqref{eqn:SNReff} assumes that the signal is constant over the $T$-stage sensing horizon and that the dependence of SNR on $\lambda_{i}(t-1)$ is linear; in the latter case, nonlinear dependences are also considered in \cite{adapEstJSTSP2013}.

After $T$ sensing stages, the full set of observations $\mbY(T)$ is used to produce an estimate $\mbxh$ of $\mbx$.  As in \cite{adapEstJSTSP2013}, we assume that the nonzero components of $\mbx$ are of primary interest and seek to minimize the mean 
$q$th-power error over the signal support:
\begin{equation}\label{eqn:MQEROI}
\E \left\{ \sum_{i=1}^{N} I_{i} \abs{\xh_{i} - x_{i}}^{q} \right\},
\end{equation}
where $q > 0$ and the expectation is taken over $\mbI$, $\mbx$, and $\mbY(T)$.  Setting $q = 2$ yields the familiar mean squared error (MSE). 
In addition to accounting for amplitude estimation error directly, the cost function \eqref{eqn:MQEROI} also indirectly promotes better performance in estimating the support of the signal, i.e., detecting nonzero signal components, as demonstrated in \cite{bashan2008,adapEstICASSP2013}.  This is because the lowest values of \eqref{eqn:MQEROI} are achieved by concentrating sensing effort on the signal support, which in turn requires reliable identification of the support. 
Missed nonzero components lead to insufficient effort allocations and higher error, while false alarms, i.e., zero-valued components mistaken as nonzero, are also penalized by \eqref{eqn:MQEROI} because they divert resources away from the true signal support.

\section{Effort allocation policies}
\label{sec:policy}

\subsection{Optimal two-stage policy}
\label{subsec:policyOpt}

In \cite{adapEstJSTSP2013}, an effort allocation policy was derived that minimizes the mean $q$th-power error \eqref{eqn:MQEROI} 
over all two-stage policies subject to the budget constraint \eqref{eqn:budget}.  This optimal two-stage policy is the subject of study in the present work and is summarized here along with related facts for later reference. 

The policy in \cite{adapEstJSTSP2013} is obtained using dynamic programming \cite{bertsekas2005} and thus depends on the observation history $\mbY(t)$ through a collection of state variables $\mbs(t) = (\mbp(t), \mbmu(t), \mbsigma^{2}(t), \Lambda(t))$, where $p_{i}(t) = \mbbP(I_{i} = 1 \mid \mbY(t))$, $\mu_{i}(t) = \E [x_{i} \mid I_{i}=1, \mbY(t)]$, and $\sigma_{i}^{2}(t) = \var( x_{i} \mid I_{i}=1, \mbY(t) )$ for $i=1,\ldots,N$, and $\Lambda(t)$ is the remaining sensing budget after $t$ stages.  At time $t = 0$, the state variables are initialized uniformly over $i$ to the prior values $p$, $\mu$, $\sigma^{2}$, and $\Lambda(0)$, and evolve according to the following relations:
\begin{subequations}\label{eqn:stateEvol}
\begin{align}
p_{i}(t+1) &= \frac{p_{i}(t) f_{1}(y_{i}(t+1))}{p_{i}(t) f_{1}(y_{i}(t+1)) + (1 - p_{i}(t)) f_{0}(y_{i}(t+1))},\label{eqn:pEvol}\\
\mu_{i}(t+1) &= \frac{\nu^{2}\mu_{i}(t) + \lambda_{i}(t) \sigma_{i}^{2}(t) y_{i}(t+1)}{\nu^{2} + \lambda_{i}(t) \sigma_{i}^{2}(t)},\label{eqn:muEvol}\\
\sigma_{i}^{2}(t+1) &= \frac{\nu^{2} \sigma_{i}^{2}(t)}{\nu^{2} + \lambda_{i}(t) \sigma_{i}^{2}(t)},\label{eqn:sigmaEvol}\\
\Lambda(t+1) &= \Lambda(t) - \sum_{i=1}^{N} \lambda_{i}(t),\label{eqn:LambdaEvol}
\end{align}
\end{subequations}
where 
\begin{subequations}\label{eqn:f0f1}
\begin{align}
f_{0}(y_{i}(t+1)) &= f(y_{i}(t+1) \mid I_{i}=0, \mbY(t)) = \phi(y_{i}(t+1); 0, \nu^{2}/\lambda_{i}(t)),\label{eqn:f0}\\
f_{1}(y_{i}(t+1)) &= f(y_{i}(t+1) \mid I_{i}=1, \mbY(t)) = \phi(y_{i}(t+1); \mu_{i}(t), \sigma_{i}^{2}(t) + \nu^{2}/\lambda_{i}(t)),\label{eqn:f1}
\end{align}
\end{subequations}
and $\phi(\cdot; \mu, \sigma^{2})$ denotes the Gaussian probability density function (PDF) with mean $\mu$ and variance $\sigma^{2}$.  
The effort parameters $\lambda_{i}(t)$ in \eqref{eqn:stateEvol} are determined by the effort allocation policy, and specifically through \eqref{eqn:J1*}, \eqref{eqn:J0*} below.
With these definitions, 
it is shown in \cite{adapEstJSTSP2013} that the conditional mean estimator $\mbxh = \mbmu(T)$ minimizes \eqref{eqn:MQEROI} 
and that the effort allocation problem for $T = 2$ can be expressed as a two-stage optimization problem,
\begin{equation}\label{eqn:J1*}
\begin{split}
J_{1}^{\ast}(\mbs(1)) = m_{q} \nu^{q} \; \min_{\mblambda(1)} \quad &\sum_{i=1}^{N} \frac{p_{i}(1)}{\left( \nu^{2}/\sigma_{i}^{2}(1) + \lambda_{i}(1) \right)^{q/2}}\\
\text{s.t.} \quad &\sum_{i=1}^{N} \lambda_{i}(1) = \Lambda(1),
\qquad \lambda_{i}(1) \geq 0 \quad \forall \; i,
\end{split}
\end{equation}
\begin{equation}\label{eqn:J0*}
\begin{split}
J_{0}^{\ast}(\mbs(0)) = \min_{\lambda} \quad &\E \left\{ J_{1}^{\ast}(\mbs(1)) \mid \mbs(0), \lambda\ones \right\}\\
\text{s.t.} \quad &0 \leq \lambda \leq \frac{\Lambda(0)}{N},
\end{split}
\end{equation}
where $m_{q} = \E[ \abs{z}^{q} ]$, $z \sim \mcN(0,1)$, is the $q$th absolute moment of a standard Gaussian random variable.  The optimal two-stage estimation error is given by $J_{0}^{\ast}(\mbs(0))$.  The second-stage optimization \eqref{eqn:J1*} over $\mblambda(1)$ 
depends on the values of the first-stage observations $\mby(1)$ through the state $\mbs(1)$.  The first-stage optimization \eqref{eqn:J0*} is then defined recursively in terms of $J_{1}^{\ast}(\mbs(1))$.  Under the assumption of a uniform prior, the first-stage allocation $\mblambda(0)$ is also uniform by symmetry, i.e., $\mblambda(0) = \lambda \ones$ where $\ones$ is a vector of ones, thus making \eqref{eqn:J0*} a one-dimensional optimization.  The expectation in \eqref{eqn:J0*} is taken over $\mby(1)$, which has i.i.d.~components with distribution 
\begin{align}
f_{p}(y_{i}(1)) &= p f_{1}(y_{i}(1)) + (1-p) f_{0}(y_{i}(1))\label{eqn:fp}\\
&= p \phi(y_{i}(1); \mu, \sigma^{2} + \nu^{2}/\lambda) + (1-p) \phi(y_{i}(1); 0, \nu^{2}/\lambda)\nonumber
\end{align}
parameterized by $\mbs(0) = \{ p, \mu, \sigma^{2}, \Lambda(0) \}$ and $\lambda$.

The optimal solution to \eqref{eqn:J1*} has an explicit form as derived in \cite{adapEstJSTSP2013}.  Define $\gamma = 2 / (q + 2)$ and $\pi$ to be an index permutation that sorts the quantities 
$p_{i}^{\gamma}(1) \sigma_{i}^{2}(1)$ in non-increasing order:
\begin{equation}\label{eqn:pi}
p_{\pi(1)}^{\gamma}(1) \sigma_{\pi(1)}^{2}(1) \geq p_{\pi(2)}^{\gamma}(1) \sigma_{\pi(2)}^{2}(1) \geq \dots \geq p_{\pi(N)}^{\gamma}(1) \sigma_{\pi(N)}^{2}(1).\end{equation}
Then the optimal solution $\mblambda^{\ast}(1)$ to \eqref{eqn:J1*} is given by 
\begin{equation}\label{eqn:lambda*}
\lambda_{\pi(i)}^{\ast}(1) = \begin{cases}
C p_{\pi(i)}^{\gamma}(1) 
- \frac{\nu^{2}}{\sigma_{\pi(i)}^{2}(1)}, & i = 1,\ldots,k,\\
0, & i = k+1,\ldots,N,
\end{cases}
\end{equation}
where
\begin{equation}\label{eqn:C}
C = \frac{\Lambda(1) + \sum_{j=1}^{k} \frac{\nu^{2}}{\sigma_{\pi(j)}^{2}(1)}}
{\sum_{j=1}^{k} p_{\pi(j)}^{\gamma}(1)}.
\end{equation}
Equation \eqref{eqn:lambda*} shows that the optimal allocations are thresholded to zero after a certain point in the rank order and that the nonzero allocations increase 
with the probabilities $p_{i}(1)$ raised to the power $\gamma$ and decrease with the precisions $1/\sigma_{i}^{2}(1)$.  The number of nonzero allocations $k$ is determined by the interval $(b(k-1), b(k)]$ in which the budget parameter $\Lambda(1)$ falls, where $b(k)$ is the monotone non-decreasing sequence  
\begin{equation}\label{eqn:b}
b(k) = \frac{\nu^{2}}{p_{\pi(k+1)}^{\gamma}(1) \sigma_{\pi(k+1)}^{2}(1)} \sum_{i=1}^{k} p_{\pi(i)}^{\gamma}(1) - \sum_{i=1}^{k} \frac{\nu^{2}}{\sigma_{\pi(i)}^{2}(1)}, \quad k = 0,\ldots,N-1,
\end{equation}
and $b(N) = \infty$.  

The optimal solution to \eqref{eqn:J0*} can therefore be computed by sweeping $\lambda$ from $0$ to $\Lambda(0)/N$, approximating the expected cost-to-go $\E \{ J_{1}^{\ast}(\mbs(1)) \}$ for each $\lambda$, for example by Monte Carlo, 
and using \eqref{eqn:pi}--\eqref{eqn:b} to compute $J_{1}^{\ast}(\mbs(1))$.  
In the course of the analysis in Section \ref{subsec:pFixed}, 
 a closed-form upper bound is derived on the expected cost-to-go, suggesting as a by-product a simplification of the optimal procedure above in which the expected cost-to-go is replaced by its upper bound and Monte Carlo simulation is avoided.  This is discussed in more detail in Section \ref{subsec:pFixed}.

\subsection{Non-adaptive and oracle policies}
\label{subsec:policyNonAdapOracle}

Our results in Section \ref{sec:results} compare the optimal two-stage policy in Section \ref{subsec:policyOpt} to two other policies: single-stage non-adaptive sensing, and an oracle policy that has full knowledge of the true signal support and distributes effort only over the support.  In this subsection, we derive the mean estimation error of the non-adaptive and oracle policies.

The non-adaptive estimation error can be determined as a special case of two-stage sensing by setting $\lambda = 0$ in \eqref{eqn:J0*}, i.e., by skipping the first observation stage.  As a consequence, the state remains unchanged from its initial value, $\mbs(1) = \mbs(0)$, and the non-adaptive estimation error $J^{\na}(\mbs(0))$ is equal to $J_{1}^{\ast}(\mbs(0))$.  By symmetry, the optimal solution to \eqref{eqn:J1*} is to divide the sensing budget $\Lambda(0)$ evenly, yielding 
\begin{equation}\label{eqn:MQEnaUnnorm}
J^{\na}(\mbs(0)) = \frac{m_{q} \nu^{q}N p}{\left( (\nu^{2} / \sigma^{2}) + (\Lambda(0) / N) \right)^{q/2}}.
\end{equation}
To express \eqref{eqn:MQEnaUnnorm} in a more interpretable form, we normalize the budget $\Lambda(0)$ so that $\Lambda(0) / N = 1$, which has the effect of scaling $\nu^{2}$ by $N/\Lambda(0)$ in the effective SNR \eqref{eqn:SNReff}. 
We also define the ratio $r \equiv \sigma^{2} / \nu^{2}$ as a measure of effective SNR.  Because of the normalization of $\Lambda(0)$, $r$ accounts not only for the intrinsic strength of the nonzero signal amplitudes relative to the noise, but also for the size of the sensing budget. 
With these definitions, \eqref{eqn:MQEnaUnnorm} can be rewritten as 
\begin{equation}\label{eqn:MQEna}
J^{\na}(\mbs(0)) = \frac{m_{q} \sigma^{q} N p}{(1 + r)^{q/2}}.
\end{equation}
The numerator can be interpreted as the mean estimation error incurred a priori, i.e., with $\mbxh = \mbmu(0)$, while the denominator is the reduction factor due to making observations $\mby(1)$. 

%
%

The oracle estimation error can be similarly determined by setting $\lambda = 0$ in \eqref{eqn:J0*} but instead assuming that the probabilities $p_{i}(1)$ collapse to $1$ for the $k$ locations with nonzero signal components and to $0$ elsewhere.  The total budget $\Lambda(0) = N$ is then divided into shares of $N/k$ for each nonzero component, resulting in an oracle error of 
\[
J^{o}(\mbs(0)) = m_{q} \nu^{q} \E \left\{ \frac{k}{\left( \nu^{2} / \sigma^{2} + N/k \right)^{q/2}} \right\}
= m_{q} \sigma^{q} \E \left\{ \frac{k}{(1 + rN/k)^{q/2}} \right\},
\]
where the expectation is over $k$, a binomial random variable with parameters $N$ and $p$.  Since the function $k / (1 + rN/k)^{q/2}$ is convex with respect to $k \geq 0$, an application of Jensen's inequality yields 
\[
J^{o}(\mbs(0)) \geq m_{q} \sigma^{q} \frac{\E\{k\}}{\left( 1 + rN/\E\{k\} \right)^{q/2}} = \frac{m_{q} \sigma^{q} N p}{(1 + r/p)^{q/2}}.
\]

We define the estimation gain of an allocation policy as its error reduction factor relative to non-adaptive uniform allocation.  The gain of the oracle policy is thus bounded as 
\begin{equation}\label{eqn:gainOracle}
G^{o} \leq \left( \frac{1 + r/p}{1 + r} \right)^{q/2}.
\end{equation}

\section{Performance guarantees}
\label{sec:results}

Previous work \cite{adapEstJSTSP2013,bashan2008} has shown empirically that the optimal two-stage policy in Section \ref{subsec:policyOpt} can achieve substantially lower estimation error than the non-adaptive strategy of uniform effort allocation.  The contribution of this paper is to provide analytical guarantees on the gain due to adaptation. 
Of particular interest is the dependence of the gain on several key quantities: the sparsity of the signal as represented by the fraction $p$, the SNR, and the sensing budget.  With regard to sparsity, two cases are distinguished: the first in Section \ref{subsec:pFixed} in which $p$ is fixed, representing a fixed fraction of nonzero components, and the second in Section \ref{subsec:pVanishing} in which $p$ decreases to zero as $N$ increases, corresponding to sublinear growth in the number of nonzero components as a function of $N$.  As in Section \ref{subsec:policyNonAdapOracle}, we normalize the total sensing budget $\Lambda(0)$ to $N$.  The SNR and the sensing budget can then be summarized by the ratio $r = \sigma^{2} / \nu^{2}$ introduced in Section \ref{subsec:policyNonAdapOracle}, which increases as either the intrinsic noise variance decreases or the sensing budget increases.  We also define $s \equiv \mu^{2} / \sigma^{2}$ to represent the prior degree of certainty in the nonzero signal amplitudes; $s$ will be regarded as a fixed parameter. 


As noted in Section \ref{sec:prob}, the improvement in the mean estimation error \eqref{eqn:MQEROI} due to adaptation is determined by the ability to concentrate sensing resources on the true signal support and thus increase the effective SNR.  This in turn depends on the detectability of nonzero signal components, which is characterized by the contrast between the signal-absent measurement likelihood $f_{0}$ \eqref{eqn:f0} and the signal-present likelihood $f_{1}$ \eqref{eqn:f1}.  
Theorem \ref{thm:pFixed} shows that the relevant measures of contrast are Chernoff coefficients between $f_{1}(y_{i}(1))$ 
%
%
and the unconditional likelihood $f_{p}(y_{i}(1))$ \eqref{eqn:fp}, which is a mixture of $f_{1}(y_{i}(1))$ and $f_{0}(y_{i}(1))$
%
%
with weights $p$ and $1-p$ respectively.  In the sequel, we use $f_{0}(y)$ and $f_{1}(y)$ as a shorthand to refer to $f_{0}(y_{i}(1))$ and $f_{1}(y_{i}(1))$, noting that these distributions do not depend on $i$ due to the assumed homogeneous distribution of amplitudes over the signal support.  
The Chernoff coefficient $C_{p}^{\gamma}$ is then defined as 
\begin{equation}\label{eqn:Cp}
C_{p}^{\gamma} = \int_{-\infty}^{\infty} f_{1}^{\gamma}(y) f_{p}^{1-\gamma}(y) dy = \int_{-\infty}^{\infty} f_{1}^{\gamma}(y) \left( p f_{1}(y) + (1-p) f_{0}(y) \right)^{1-\gamma} dy,
\end{equation}
where the exponent $\gamma = 2/(q+2) \in (0, 1)$ as before and is determined by the exponent $q$ chosen to characterize the loss function \eqref{eqn:MQEROI}.  In general, Chernoff coefficients measure the overlap between two probability distributions and take values between $0$ and $1$, with $0$ corresponding to no overlap and $1$ to perfect overlap.  For the special case of MSE ($q = 2$), $\gamma = 1/2$ and the Chernoff coefficient reduces to the Bhattacharyya coefficient 
%
\[
C_{p}^{1/2 }= \int_{-\infty}^{\infty} \sqrt{f_{1}(y) f_{p}(y)} dy, 
\]
%
which is related to the Hellinger distance $H(f_{1}, f_{p})$ between two distributions via $H(f_{1},f_{p}) = \sqrt{1 - C_{p}^{1/2}}$.

\subsection{Estimation gain:  large $N$ and fixed fraction $p$ of nonzero signal components}
\label{subsec:pFixed}

We consider first the case in which the sparsity parameter $p \in [0,1]$ is fixed and establish the following bounds on the performance of two-stage adaptive sensing.  Similar to \cite{haupt2011,malloy2011b,tajer2012,iwen2012,haupt2012,malloy2012}, we focus more attention on the limit as the signal dimension $N \to \infty$, which simplifies the form of the bounds. 
\begin{theorem}\label{thm:pFixed}
Assume that the signal $\mbx$ and observations $\mby(1), \mby(2)$ follow the model in Section \ref{sec:prob} with sensing budget $\Lambda(0) = N$.  Then for any $\epsilon > 0$, the optimal two-stage estimation error $J_{0}^{\ast}(\mbs(0))$ is bounded from above as 
\begin{equation}\label{eqn:J0*UBp}
J_{0}^{\ast}(\mbs(0)) \leq \frac{m_{q} \sigma^{q} N p}{\left[ 1 + r + r \displaystyle\max_{\lambda\in[0,1]} \left( (1+\epsilon)^{-1} \left(C_{p}^{\gamma}\right)^{-\frac{1}{1-\gamma}} - 1 \right)(1 - \lambda) \right]^{q/2}}
\end{equation}
with probability at least 
\[
\begin{cases}
1 - \exp \left( -\dfrac{\left(C_{p}^{\gamma}\right)^{2} N p^{\gamma} \epsilon^{2}}{2 \left(p^{\gamma} \left(C_{p}^{2\gamma} - \left(C_{p}^{\gamma}\right)^{2}\right) + \epsilon C_{p}^{\gamma} /3\right)} \right), & 0 < \gamma \leq \frac{1}{2},\\
1 - \exp \left( -\dfrac{\left(C_{p}^{\gamma}\right)^{2} N p^{\gamma} \epsilon^{2}}{2 \left(p^{1-\gamma} - p^{\gamma} \left(C_{p}^{\gamma}\right)^{2} + \epsilon C_{p}^{\gamma} /3\right)} \right), & \frac{1}{2} < \gamma < 1.
\end{cases}
\]
%
In the limit as the signal dimension $N \to \infty$ and with $\epsilon = \omega\left(1/\sqrt{N}\right)$, we have 
\begin{equation}\label{eqn:J0*UBpInf}
\lim_{N\to\infty} \frac{1}{m_{q} \sigma^{q} N p} J_{0}^{\ast}(\mbs(0)) \leq \frac{1}{\left[ 1 + r + r \displaystyle\max_{\lambda\in[0,1]} \left( \left(C_{p}^{\gamma}\right)^{-\frac{1}{1-\gamma}} - 1 \right)(1 - \lambda) \right]^{q/2}}
\end{equation}
and a corresponding bound on the optimal two-stage gain $G = J^{\na}(\mbs(0)) / J_{0}^{\ast}(\mbs(0))$:
\begin{equation}\label{eqn:Gp}
\lim_{N\to\infty} G \geq \left[ 1 + \frac{r}{r+1} \max_{\lambda\in [0,1]} \left( \left(C_{p}^{\gamma}\right)^{-\frac{1}{1-\gamma}} - 1 \right)(1 - \lambda) \right]^{q/2}. 
\end{equation}
Equality holds asymptotically in \eqref{eqn:J0*UBpInf}, \eqref{eqn:Gp} in the limits $r \to 0$ and $r \to \infty$ with $\lambda \to 0$ simultaneously in the latter case. 
\end{theorem}
%
\begin{IEEEproof}
First we bound the cost-to-go $J_{1}^{\ast}(\mbs(1))$ in \eqref{eqn:J1*}.  Although the optimal solution to \eqref{eqn:J1*} is given explicitly in \eqref{eqn:pi}--\eqref{eqn:b}, it requires determining the number of nonzero allocations $k$ using \eqref{eqn:b}, which depends in turn on the rank order of the random variables $\sqrt{p_{i}(1)}$.  
Since a suboptimal allocation can only increase the cost-to-go, we can upper bound $J_{1}^{\ast}(\mbs(1))$ by evaluating \eqref{eqn:J1*} with the specific non-sparse allocation: 
\begin{equation}\label{eqn:lambdaSubopt}
\lambda_{i}(1) = \Lambda(1) \frac{p_{i}^{\gamma}(1)}{\sum_{j=1}^{N} p_{j}^{\gamma}(1)}.
\end{equation}
The allocation \eqref{eqn:lambdaSubopt} preserves the 
$\gamma$-power dependence on the probabilities $p_{i}(1)$ as in the optimal allocation \eqref{eqn:lambda*} but does not threshold any of the $\lambda_{i}(1)$ to zero.  A particular case of \eqref{eqn:lambdaSubopt} with $\gamma = 1/2$ was first proposed in \cite{bashan2008} and was shown empirically to result in only a small performance loss relative to the optimal allocation. 

Substituting \eqref{eqn:lambdaSubopt} into \eqref{eqn:J1*}, we obtain 
\begin{align}
J_{1}^{\ast}(\mbs(1)) &\leq m_{q} \nu^{q} \sum_{i=1}^{N} \frac{p_{i}(1)}{\left( \nu^{2}/\sigma_{i}^{2}(1) + \Lambda(1) p_{i}^{\gamma}(1) / \sum_{j=1}^{N} p_{j}^{\gamma}(1) \right)^{q/2}}\nonumber\\
&= m_{q} \nu^{q} \sum_{i=1}^{N} \frac{p_{i}(1)}{\left( \nu^{2}/\sigma^{2} + \lambda + N(1-\lambda) p_{i}^{\gamma}(1) / \sum_{j=1}^{N} p_{j}^{\gamma}(1) \right)^{q/2}}\label{eqn:J1*UB1}\\
&= m_{q} \sigma^{q} \sum_{i=1}^{N} \frac{p_{i}(1)}{\left( 1 + r\lambda + r(1-\lambda) p_{i}^{\gamma}(1) \left( \frac{1}{N} \sum_{j=1}^{N} p_{j}^{\gamma}(1) \right)^{-1} \right)^{q/2}}.\label{eqn:J1*UB2}
\end{align} 
In the first equality \eqref{eqn:J1*UB1}, we have used \eqref{eqn:sigmaEvol} and \eqref{eqn:LambdaEvol}, $\lambda_{i}(0) = \lambda$, and the normalization $\Lambda(0) = N$, while in the second equality \eqref{eqn:J1*UB2}, we have used the definition of $r$ from Section \ref{subsec:policyNonAdapOracle}.  The bound in \eqref{eqn:J1*UB2} is monotone non-decreasing in the quantity $\frac{1}{N} \sum_{j=1}^{N} p_{j}^{\gamma}(1)$.  Hence an application of Lemma \ref{lem:psqrt} in Appendix \ref{app:tailBounds} yields 
\begin{equation}\label{eqn:J1*UB3}
J_{1}^{\ast}(\mbs(1)) \leq m_{q} \sigma^{q} \sum_{i=1}^{N} \frac{p_{i}(1)}{\left( 1 + r\lambda + r(1-\lambda) p_{i}^{\gamma}(1) \left( (1+\epsilon) p^{\gamma} C_{p}^{\gamma} \right)^{-1} \right)^{q/2}}
\end{equation}
with probabilities as indicated in Lemma \ref{lem:psqrt} and the theorem statement. 

Next we bound the two-stage cost $J_{0}^{\ast}(\mbs(0))$ by combining \eqref{eqn:J0*} and \eqref{eqn:J1*UB3} to yield 
\begin{align}
J_{0}^{\ast}(\mbs(0)) &\leq m_{q} \sigma^{q} \min_{\lambda\in[0,1]} \; \E \left\{ \sum_{i=1}^{N} \frac{p_{i}(1)}{\left( 1 + r\lambda + r(1-\lambda) p_{i}^{\gamma}(1) \left( (1+\epsilon) p^{\gamma} C_{p}^{\gamma} \right)^{-1} \right)^{q/2}} \right\}\nonumber\\
&= m_{q} \sigma^{q} N \min_{\lambda\in[0,1]} \; \E \left\{ \frac{p_{i}(1)}{\left( 1 + r\lambda + r(1-\lambda) p_{i}^{\gamma}(1) \left( (1+\epsilon) p^{\gamma} C_{p}^{\gamma} \right)^{-1} \right)^{q/2}} \right\}\label{eqn:J0*UB1},
\end{align}
since $\left\{ p_{i}^{\gamma}(1) \right\}_{i=1}^{N}$ are identically distributed. 
The right-hand side of \eqref{eqn:J0*UB1} can be further simplified.  Using \eqref{eqn:p1}, the expectation in \eqref{eqn:J0*UB1}, which is taken with respect to the unconditional density $f_{p}(y_{i}(1))$, can be converted into an expectation with respect to $f_{1}(y_{i}(1))$, i.e., conditioned on $I_{i} = 1$.  Thus \eqref{eqn:J0*UB1} is equivalent to 
%
\begin{align*}
J_{0}^{\ast}(\mbs(0)) &\leq m_{q} \sigma^{q} N \min_{\lambda\in[0,1]} \; \E \left\{ \frac{p}{\left( 1 + r\lambda + r(1-\lambda) p_{i}^{\gamma}(1) \left( (1+\epsilon) p^{\gamma} C_{p}^{\gamma} \right)^{-1} \right)^{q/2}} \mid I_{i} = 1 \right\}\\
&= m_{q} \sigma^{q} N p \min_{\lambda\in[0,1]} \; \E \left\{ \frac{1}{\left( 1 + r\lambda + r(1-\lambda) \left(p_{i}^{-(1-\gamma)}(1)\right)^{-2/q} \left( (1+\epsilon) p^{\gamma} C_{p}^{\gamma} \right)^{-1} \right)^{q/2}} \mid I_{i} = 1 \right\}
\end{align*}
using the relations $\gamma = \frac{2}{q} \frac{q}{q+2} = \frac{2}{q} (1-\gamma)$. 
The quantity inside the expectation is of the form 
\begin{equation}\label{eqn:concaveFun}
\frac{1}{(\alpha + \beta x^{-2/q})^{q/2}}, \quad \alpha, \beta \geq 0,
\end{equation}
%
which can be shown to be a concave function of $x$ by differentiating twice.  (More informally, it can be seen that the function transitions from linear for small $x$ to constant for large $x$.)  Applying Jensen's inequality with $x = p_{i}^{-(1-\gamma)}(1)$ yields  
\begin{align}
J_{0}^{\ast}(\mbs(0)) &\leq m_{q} \sigma^{q} N p \min_{\lambda\in[0,1]} \frac{1}{\left( 1 + r\lambda + r(1-\lambda) \left( \E\left[ p_{i}^{-(1-\gamma)}(1) \mid I_{i} = 1 \right] \right)^{-2/q} \left( (1+\epsilon) p^{\gamma} C_{p}^{\gamma} \right)^{-1} \right)^{q/2}}\nonumber\\
&= \frac{m_{q} \sigma^{q} N p}{\left( \displaystyle\max_{\lambda\in[0,1]} 1 + r\lambda + r(1-\lambda) \left( p^{-(1-\gamma)} C_{p}^{\gamma} \right)^{-2/q} \left( (1+\epsilon) p^{\gamma} C_{p}^{\gamma} \right)^{-1} \right)^{q/2}}\nonumber\\
&= \frac{m_{q} \sigma^{q} N p}{\left( \displaystyle\max_{\lambda\in[0,1]} 1 + r\lambda + r(1-\lambda) (1+\epsilon)^{-1} \left(C_{p}^{\gamma}\right)^{-\frac{1}{1-\gamma}} \right)^{q/2}}.\label{eqn:J0*UB2}
\end{align}
In the first equality above, we have used \eqref{eqn:p1} to obtain 
\[
\E \left[ p_{i}^{-(1-\gamma)}(1) \mid I_{i} = 1 \right] = \int_{-\infty}^{\infty} p^{-(1-\gamma)} \left( \frac{f_{p}(y)}{f_{1}(y)} \right)^{1-\gamma} f_{1}(y) dy = p^{-(1-\gamma)} C_{p}^{\gamma}.
\]

The non-asymptotic bound \eqref{eqn:J0*UBp} follows from a rearrangement of the denominator in \eqref{eqn:J0*UB2}.  Taking $\epsilon$ to $0$ at a rate $\omega\left(1/\sqrt{N}\right)$ yields the asymptotic bound \eqref{eqn:J0*UBpInf} with probability converging to $1$.  
The bound \eqref{eqn:Gp} on the gain $G$ follows from a comparison with the non-adaptive estimation error \eqref{eqn:MQEna}.  

In the infinite-dimensional regime $N \to \infty$, the two approximations made in the above proof are the use of the suboptimal second-stage allocation in \eqref{eqn:lambdaSubopt} and Jensen's inequality applied to the function in \eqref{eqn:concaveFun}.  Both of these approximations are tight in the limits $r \to 0$ and $r \to \infty$, $\lambda \to 0$, and hence the asymptotic bounds \eqref{eqn:J0*UBpInf}, \eqref{eqn:Gp} are also tight.  To show that the suboptimal allocation \eqref{eqn:lambdaSubopt} can be optimal, we use the relation $\nu^{2} / \sigma_{i}^{2}(1) = 1/r + \lambda$ (cf.~\eqref{eqn:J1*UB1}) to express the nonzero optimal allocations in \eqref{eqn:lambda*} as 
\begin{align}
\lambda_{\pi(i)}^{\ast}(1) &= \left( \Lambda(1) + k \left( \frac{1}{r} + \lambda \right) \right) \frac{p_{\pi(i)}^{\gamma}(1)}{\sum_{j=1}^{k} p_{\pi(j)}^{\gamma}(1)} - \left( \frac{1}{r} + \lambda \right)\nonumber\\
&= \Lambda(1) \frac{p_{\pi(i)}^{\gamma}(1)}{\sum_{j=1}^{k} p_{\pi(j)}^{\gamma}(1)} + \left( \frac{k p_{\pi(i)}^{\gamma}(1)}{\sum_{j=1}^{k} p_{\pi(j)}^{\gamma}(1)} - 1 \right) \left( \frac{1}{r} + \lambda \right).\label{eqn:lambda*2}
\end{align}
Similarly, \eqref{eqn:b} can be rewritten as 
\begin{equation}\label{eqn:b2}
b(k) = \left( \frac{\sum_{i=1}^{k} p_{\pi(i)}^{\gamma}(1)}{p_{\pi(k+1)}^{\gamma}(1)} - k \right) \left( \frac{1}{r} + \lambda \right).
\end{equation}
As $r \to 0$, i.e., for vanishing SNR or sensing budget, the probabilities $p_{i}(1)$ deviate less and less from their common prior value of $p$.  Thus the quantity in the first set of parentheses vanishes in both \eqref{eqn:lambda*2} and \eqref{eqn:b2}.  It follows that $b(N-1) = 0 < \Lambda(1)$, the number of nonzero allocations $k = N$, and \eqref{eqn:lambda*2} coincides with the suboptimal allocation \eqref{eqn:lambdaSubopt}.  For $r \to \infty$ and $\lambda \to 0$, the quantity in the second set of parentheses vanishes in \eqref{eqn:lambda*2} and \eqref{eqn:b2} and thus \eqref{eqn:J0*UBpInf}, \eqref{eqn:Gp} are tight.

As for Jensen's inequality, equality is achieved if the function in \eqref{eqn:concaveFun} degenerates into an affine function of $x$.  This is indeed the case for $r \to 0$ since $\beta$ in \eqref{eqn:concaveFun} converges to $0$ and \eqref{eqn:concaveFun} becomes a constant function.  For $r \to \infty$ and $\lambda \to 0$, $\alpha = 1 + r\lambda$ becomes negligible compared to $\beta = \Theta(r)$ and \eqref{eqn:concaveFun} becomes linear. 
\end{IEEEproof}

\begin{remark}
While Theorem \ref{thm:pFixed} is stated for two-stage policies, the bounds also hold for the OLFC policies in \cite{adapEstJSTSP2013} with more than two stages because these policies improve monotonically as the number of stages increases under a fixed total sensing budget (see \cite[Prop.~2]{adapEstJSTSP2013}).  However, the bounds may not be tight for $T > 2$.  The same comments apply to performance bounds to be presented in the sequel.  As discussed briefly in Section \ref{sec:concl}, the derivation of more refined bounds for $T > 2$, i.e., a sequence of tight bounds for increasing $T$, is a subject for future work.
\end{remark}

The bound in \eqref{eqn:Gp} guarantees that the gain $G$ is greater than $1$ for $r > 0$ since the Chernoff coefficient $C_{p}^{\gamma}$ between $f_{1}$ and $f_{p}$ is strictly less than $1$ for $r > 0$.  Further analysis of $J_{0}^{\ast}(\mbs(0))$ and $G$ requires computing $C_{p}^{\gamma}$,
but unfortunately, a closed-form expression 
is not available when $p \in (0, 1)$.  As a first step, we evaluate the bounds 
in the asymptotic limits of low and high SNR/sensing budget.  In the former case ($\nu^{2} \to \infty$, $r \to 0$), 
it can be seen from \eqref{eqn:f0f1} that the distributions $f_{0}(y_{i}(1))$ and $f_{1}(y_{i}(1))$ become more diffuse and similar to each other.  Therefore the 
Chernoff coefficient $C_{p}^{\gamma}$ approaches $1$ as $r \to 0$, and so too does the gain $G$. 
For high SNR/sensing budgets ($\nu^{2} \to 0$, $r \to \infty$), $f_{0}$ and $f_{1}$ become increasingly concentrated and their overlap tends to zero.  Hence 
\eqref{eqn:Cp} implies that $C_{p}^{\gamma}$ approaches $p^{1-\gamma}$ and $(C_{p}^{\gamma})^{-\frac{1}{1-\gamma}}$ approaches $1/p$.  If we take $\lambda \to 0$ at a slower rate than $\nu^{2}$ so that $\nu^{2} / \lambda$ still approaches zero, then it can be seen from \eqref{eqn:Gp} 
that $G$ tends to $(1/p)^{q/2}$ as $r \to \infty$.  
This is the same gain as that of the oracle \eqref{eqn:gainOracle} for $r \to \infty$, thus confirming analytically that two-stage adaptive sensing can approach oracle performance at high SNR/sensing budget, as observed previously in \cite{adapEstJSTSP2013,bashan2008}.

More 
insight can be obtained by bounding $C_{p}^{\gamma}$ in terms of $C_{0}^{\gamma}$, the Chernoff coefficient between $f_{1}$ and $f_{0}$.  The latter coefficient can be computed in closed form since it involves the two Gaussian distributions specified in \eqref{eqn:f0f1}.  From the general formula for two Gaussians with parameters $\mu_{0}, \sigma_{0}^{2}$ and $\mu_{1}, \sigma_{1}^{2}$, we have 
\begin{align}
C_{0}^{\gamma} &= \sqrt{\frac{\sigma_{0}^{2\gamma} \sigma_{1}^{2(1-\gamma)}}{\gamma \sigma_{0}^{2} + (1-\gamma) \sigma_{1}^{2}}} \exp\left( -\frac{\gamma (1-\gamma) (\mu_{1} - \mu_{0})^{2}}{2 (\gamma \sigma_{0}^{2} + (1-\gamma) \sigma_{1}^{2})} \right)\nonumber\\
&= \sqrt{\frac{\left(\nu^{2} / \lambda\right)^{\gamma} \left(\sigma^{2} + \nu^{2} / \lambda\right)^{1-\gamma}}{\nu^{2} / \lambda + (1-\gamma) \sigma^{2}}} \exp\left( -\frac{\gamma (1-\gamma) \mu^{2}}{2 (\nu^{2} / \lambda + (1-\gamma) \sigma^{2})} \right)\nonumber\\
&= \sqrt{\frac{\left(1 + r\lambda\right)^{1-\gamma}}{1 + (1-\gamma) r\lambda}} \exp\left( -\frac{\gamma (1-\gamma) sr\lambda}{2 (1 + (1-\gamma) r\lambda)} \right),\label{eqn:C0}
\end{align}
using the definitions of $r$ and $s$.  The following proposition provides upper bounds on $C_{p}^{\gamma}$ in terms of $C_{0}^{\gamma}$ and is proved in Appendix \ref{app:pfCp}. 
Substituting the upper bounds in place of $C_{p}^{\gamma}$ in Theorem \ref{thm:pFixed} yields bounds on the two-stage estimation error $J_{0}^{\ast}(\mbs(0))$ and gain $G$ that are weaker than before but more easily computed. 
The first bound \eqref{eqn:CpUB1} involves the standard Gaussian cumulative distribution function (CDF) while the second bound \eqref{eqn:CpUB2} is weaker but does not require computing the Gaussian CDF. 
\begin{prop}\label{prop:Cp}
For any $p \in [0, 1]$ and $\gamma \in (0, 1)$, the Chernoff coefficient $C_{p}^{\gamma}$ \eqref{eqn:Cp} is bounded in terms of $C_{0}^{\gamma}$ \eqref{eqn:C0} as follows:
\begin{subequations}
\begin{align}
C_{p}^{\gamma} &\leq \min \left\{ p^{1-\gamma} \left(1 - \gamma + \gamma \mbbP_{1}(\mcY_{1})\right) + (1-p)^{1-\gamma} \left(\mbbP_{01}(\mcY_{0}) C_{0}^{\gamma} + (1-\gamma) \left(\frac{1-p}{p}\right)^{\gamma} \mbbP_{0}(\mcY_{1}) \right), 1 \right\}\label{eqn:CpUB1}\\
&\leq \min \left\{ p^{1-\gamma} + (1-p)^{1-\gamma} C_{0}^{\gamma}, 1 \right\},\label{eqn:CpUB2}
\end{align}
\end{subequations}
where
\begin{align*}
\mcY_{0} &= \{ y : p f_{1}(y) < (1-p) f_{0}(y) \},\\
\mcY_{1} &= \{ y : p f_{1}(y) > (1-p) f_{0}(y) \},
\end{align*}
$\mbbP_{0}$, $\mbbP_{1}$ denote probability under $f_{0}$, $f_{1}$, and $\mbbP_{01}$ denotes probability under 
\[
f_{01}(y) = \mcN\left(y; \frac{\gamma\mu}{1 + (1-\gamma) r\lambda}, \frac{1 + r\lambda}{1 + (1-\gamma) r\lambda} \frac{\nu^{2}}{\lambda} \right).
\]
The bounds are asymptotically tight in the limit $r\lambda \to \infty$. 
Detailed expressions for $\mbbP_{1}(\mcY_{1})$, $\mbbP_{01}(\mcY_{0})$, and $\mbbP_{0}(\mcY_{1})$ are given below in terms of the standard Gaussian CDF $\Phi$:
\begin{subequations}
\begin{align}
\mbbP_{1}(\mcY_{1}) &= \Phi(z_{1}^{+}) + \Phi(z_{1}^{-}),\label{eqn:P1}\\
\mbbP_{01}(\mcY_{0}) &= \Phi(z_{01}^{+}) - \Phi(z_{01}^{-}),\label{eqn:P01}\\
\mbbP_{0}(\mcY_{1}) &= \Phi(z_{0}^{+}) + \Phi(z_{0}^{-}),\label{eqn:P0}\\
z_{1}^{\pm} &= \sqrt{\frac{s}{r\lambda}} \left( \pm\sqrt{1 + r\lambda} - \sqrt{1 + \frac{1}{s} (2\eta + \log(1 + r\lambda))} \right),\label{eqn:z1}\\
z_{01}^{\pm} &= \sqrt{\frac{s(1 + (1-\gamma) r\lambda)}{r\lambda}} \left( -\frac{\sqrt{1 + r\lambda}}{1 + (1-\gamma) r\lambda} \pm \sqrt{1 + \frac{1}{s} (2\eta + \log(1 + r\lambda))} \right),\label{eqn:z01}\\
z_{0}^{\pm} &= \sqrt{\frac{s}{r\lambda}} \left( \pm 1 - \sqrt{(1 + r\lambda) \left(1 + \frac{1}{s} (2\eta + \log(1 + r\lambda) )\right)} \right),\label{eqn:z0}
\end{align}
\end{subequations}
where $\eta = \log\frac{1-p}{p}$.
\end{prop}

For the special case of $\gamma = 1/2$ corresponding to the Bhattacharyya coefficient and MSE, an alternative upper bound can be derived that is tighter for small $r\lambda$ as demonstrated in Appendix \ref{app:CpBCp}.  As explained in the proof in Appendix \ref{app:pfCp}, it does not seem straightforward to generalize this bound to arbitrary $\gamma$. 
\begin{prop}\label{prop:BCp}
For any $p \in [0, 1]$, the Bhattacharrya coefficient 
$C_{p}^{1/2}$ is bounded in terms of 
$C_{0}^{1/2}$ as follows: 
\[
C_{p}^{1/2} \leq \min\left\{ \frac{p}{\sqrt{p} + \sqrt{1-p}} + \frac{\sqrt{p(1-p)}}{\sqrt{p} + \sqrt{1-p}} \mbbP_{1}(\mcY'_{1}) + 
C_{0}^{1/2} \left( \frac{1-p}{\sqrt{p} + \sqrt{1-p}} + \frac{\sqrt{p(1-p)}}{\sqrt{p} + \sqrt{1-p}} \mbbP_{01}(\mcY'_{0}) \right), 1 \right\},
\]
where  
\begin{align*}
\mcY'_{0} &= \{ y : f_{1}(y) < f_{0}(y) \},\\
\mcY'_{1} &= \{ y : f_{1}(y) > f_{0}(y) \},
\end{align*}
%
%
%
and $\mbbP_{1}(\mcY'_{1})$, $\mbbP_{01}(\mcY'_{0})$ are given by the right-hand sides of \eqref{eqn:P1}, \eqref{eqn:P01} but with $\eta = 0$ and $\gamma = 1/2$ in \eqref{eqn:z1} and \eqref{eqn:z01}.
\end{prop}
%

%
%

Propositions \ref{prop:Cp} and \ref{prop:BCp} provide approximations to the Chernoff coefficient $C_{p}^{\gamma}$ that simplify the evaluation of the bounds 
in Theorem \ref{thm:pFixed}. 
We use these approximations in Section \ref{sec:numVal} to compute lower bounds on the gain $G$.  The combination of Theorem \ref{thm:pFixed} with either Propositions \ref{prop:Cp} or \ref{prop:BCp} also offers a simpler way of determining the fraction $\lambda$ of the sensing budget to be allocated to the first stage.  Instead of optimizing \eqref{eqn:J0*} directly, which requires 
exact evaluation of or at least an accurate approximation to the expected cost-to-go $\E \left\{ J_{1}^{\ast}(\mbs(1)) \right\}$, we optimize the asymptotic upper bound on $\E \left\{ J_{1}^{\ast}(\mbs(1)) \right\}$ in \eqref{eqn:J0*UBpInf}, which is straightforward to compute with the aid of Propositions \ref{prop:Cp} or \ref{prop:BCp}.  The value of $\lambda$ that optimizes the bound is taken to be the first-stage allocation.  In Section \ref{sec:numVal}, we compare this simplified method to the exact method \eqref{eqn:J0*} and show that the performance loss is negligible. 

The approximation to $C_{p}^{\gamma}$ in Proposition \ref{prop:Cp} becomes increasingly accurate as $r\lambda \to \infty$.  Proposition \ref{prop:Cp} can therefore be used to more quantitatively reexamine the 
limit of high SNR/sensing budget considered earlier.  
By expanding the approximation to lowest order in $1/(r\lambda)$, it is possible not only to confirm that the optimal two-stage gain converges to the oracle gain as discussed earlier, but also to determine the rate of convergence and the first-stage fraction $\lambda$ 
that asymptotically achieves the optimal gain.  The following theorem summarizes these results and a detailed proof appears in Appendix \ref{app:pfpFixedHighSNR}. 
%
\begin{theorem}\label{thm:pFixedHighSNR}
In the limit as $N \to \infty$, 
the optimal two-stage gain converges to the asymptotic oracle gain of $(1/p)^{q/2}$ at the rate below:
\[
\lim_{N\to\infty} G = 
\left(\frac{1}{p}\right)^{q/2} \left(1 - 
A_{1}(q)
\frac{(1-p)^{\frac{q+1}{q+3}}}{p^{\frac{q}{q+3}}} e^{-\frac{s}{q+3}} r^{-\frac{1}{q+3}} + O\left(r^{-\frac{1+\min\{q/2,1\}}{q+3}} \sqrt{\log r} \right) \right),
\]
where $A_{1}(q)$ is a constant that depends only on the exponent $q$:
\[
A_{1}(q) = \frac{(q+3) (q+2)^{\frac{q+2}{2(q+3)}}}{2 (q)^{\frac{q}{2(q+3)}}}.
\]
To achieve this rate, the required fraction $\lambda^{\ast}$ of the sensing budget allocated to the first stage is 
\begin{equation}\label{eqn:lambda*HighSNR}
\lambda^{\ast} = 
 \frac{A_{2}(q)}{p^{\frac{q}{q+3}} (1-p)^{\frac{2}{q+3}}} e^{-\frac{s}{q+3}} r^{-\frac{1}{q+3}},
\end{equation}
\[
A_{2}(q) = \frac{(q+2)^{\frac{q+2}{2(q+3)}}}{q^{\frac{3(q+2)}{2(q+3)}}}.
\]
\end{theorem}

Theorem \ref{thm:pFixedHighSNR} shows that the two-stage gain $G$ approaches the high-SNR oracle gain $(1/p)^{q/2}$ at a rate of $r^{-\frac{1}{q+3}}$ as $r \to \infty$, and also shows that the first-stage allocation $\lambda^{\ast}$ should decrease to zero at the same rate.  These rates quantify 
empirical findings from \cite{adapEstJSTSP2013,bashan2008}.  Moreover, the result agrees with an intuitive view of two-stage sensing as the combination of an exploration stage aimed at detecting nonzero signal components followed by an exploitation stage devoted to concentrating measurements and reducing estimation error.  Note that the ideal gain of $(1/p)^{q/2}$ can only be achieved if nearly the entire budget is allocated to the selective second stage.  As the SNR or sensing budget increases, it becomes possible to 
use a smaller fraction of the budget in the first stage while still maintaining adequate detection performance, thus reserving the bulk of the budget for 
the second stage. 
The theorem makes precise how small the first-stage fraction $\lambda$ can be, or equivalently, what constitutes sufficient detection performance to enable full exploitation in the second stage.
In the limit, the first-stage fraction decreases to zero and the concentration gain in the second stage approaches $1/p$. 

The asymptotically vanishing fraction $\lambda^{\ast} \propto r^{-\frac{1}{q+3}}$ devoted to exploration is analogous to the notion of sublinear regret for multi-arm bandits \cite{auer2002,audibert2007}.  Given this context, one may wonder why the regret cannot be made logarithmic as in many bandit problems, corresponding to a fraction of $\log(r) / r$.  Such a $\log r$ vs.~$r$ allocation between stages has also been found optimal for two-stage variable selection and predictor design \cite{firouzi2013}.  The reason is that in the present setting, the problem of detecting nonzero components is a composite hypothesis test because of uncertainty in the nonzero amplitudes, as represented by the nonzero prior variance $\sigma^{2}$.  As a consequence, the probability of detection error decreases only as a weak power of the SNR/sensing budget $r$ and not exponentially as might be expected otherwise.  To be more specific, \eqref{eqn:P1HighSNR} shows that $\mbbP_{1}(\mcY_{0})$, which is the probability of missing a nonzero component, decays as $O\left( \sqrt{\frac{\log(r\lambda)}{r\lambda}} \right)$ while \eqref{eqn:P0HighSNR} shows that the false alarm probability $\mbbP_{0}(\mcY_{1})$ decays as $O\left( \sqrt{\frac{1}{r\lambda \log(r\lambda)}} \right)$.  Thus the fraction of the budget allocated to the first stage must be significantly larger than $\log(r) / r$ to control the detection errors from the first stage and limit their effect on the exploitation stage.  The exact dependence of the first-stage fraction on $r$ is determined by the interplay between detection error and the $q$th-power estimation error \eqref{eqn:MQEROI}. 

It is also interesting to interpret Theorem \ref{thm:pFixedHighSNR} in terms of MMSE dimension, a measure of information and estimability developed in \cite{wu2011}.  Both pertain to the high-SNR regime, with MMSE dimension defined as the product of SNR and the MMSE of estimating a random variable in additive noise, in the limit of increasing SNR.  We give a brief account of the connection to MMSE dimension and refer the reader to \cite{wu2011} for an in-depth discussion.  First, it can be shown that the cost function \eqref{eqn:MQEROI} adopted herein with $q = 2$, i.e., the MSE over the signal support, is equal to the MSE in estimating $\mbx$ over all dimensions but conditioned on the support $\mbI$.  From this equivalence, the MMSE dimension for the cost function \eqref{eqn:MQEROI} is determined to be $p$, being equal to the mass of the continuous part of the distribution of each $x_{i}$ \cite[Thm.~15]{wu2011}, and unchanged when conditioned on a discrete random variable \cite[Thm.~9]{wu2011}.  This result agrees with the MSE under non-adaptive sensing \eqref{eqn:MQEna} (normalized by the dimension of $\mbx$ and the variance $\sigma^{2}$), with $r$ playing the role of SNR and taking the limit as $r \to \infty$.  On the other hand, for adaptive sensing, Theorem~\ref{thm:pFixedHighSNR} seems to imply that the MMSE dimension is smaller by an additional factor of $p$.  The discrepancy lies in two differing definitions of SNR: the first associated with $r$, which corresponds to the definition used in \cite{wu2011}, and the second being the effective SNR \eqref{eqn:SNReff}, which approaches $r/p$ due to resource concentration as $r \to \infty$.  Under the second definition, Theorem~\ref{thm:pFixedHighSNR} is indeed consistent with MMSE dimension, thus offering an independent confirmation of the SNR-boosting effect of adaptive sensing.

\subsection{Estimation gain: large $N$ and vanishing fraction $p$ of nonzero signal components}
\label{subsec:pVanishing}

We now consider the case in which the average number of nonzero signal components grows sublinearly with $N$ so that the non-sparsity fraction $p$ decreases to zero.  More precisely, it is assumed that $Np \to \infty$ as $N \to \infty$ while $p \to 0$.  As seen in Theorem \ref{thm:pVanishing} below, the main difference compared to the previous regime of fixed $p$ is that the adaptation gain is no longer limited to a finite value.  This reflects the fact that the sensing budget increases linearly with $N$ while the number of nonzero components increases more slowly.  Indeed, for the oracle policy the gain becomes infinite as $p \to 0$. 
For a non-oracle policy, 
the gain is limited only by the detectability of nonzero components, which is a function of SNR.  


For any finite $N$ and $p$, Theorem \ref{thm:pFixed} gives a probabilistic upper bound on the optimal two-stage estimation error $J_{0}^{\ast}(\mbs(0))$.  As $Np \to \infty$ and $p \to 0$, one would expect the asymptotic bounds \eqref{eqn:J0*UBpInf} and \eqref{eqn:Gp} to remain valid with the Chernoff coefficient $C_{p}^{\gamma}$ replaced by $C_{0}^{\gamma}$.  This is indeed the case but the vanishing of $p$ requires a modification to the rate at which $\epsilon$ can converge to zero while still ensuring that the probabilities in Theorem \ref{thm:pFixed} converge to $1$.  The following corollary makes these convergence rates precise. 
\begin{corollary}\label{cor:pVanishing}
Assume that the average number of nonzero signal components grows sublinearly with $N$, i.e., $Np \to \infty$ and $p \to 0$ as $N \to \infty$.  
%
%
%
%
Then with $\epsilon$ in Theorem \ref{thm:pFixed} decreasing to zero as
\[
\epsilon = \begin{cases}
\omega\left( \frac{1}{\sqrt{N}} \right), & 0 < \gamma \leq \frac{1}{2},\\
\omega\left( \frac{1}{\sqrt{N} p^{\gamma - \frac{1}{2}}} \right), & \frac{1}{2} < \gamma < 1,
\end{cases}
\]
the optimal two-stage gain $G$ is asymptotically bounded as 
\begin{equation}\label{eqn:G0}
\lim_{\substack{Np\to\infty\\p\to0}} G \geq 
\left[ 1 + \frac{r}{r+1} \max_{\lambda\in [0,1]} \left( \left(C_{0}^{\gamma}\right)^{-\frac{1}{1-\gamma}} - 1 \right)(1 - \lambda) \right]^{q/2}.
\end{equation}
\end{corollary}
\begin{IEEEproof}
For the case $0 < \gamma \leq 1/2$, suppose first that $\epsilon = \omega\left( p^{\gamma} \right)$.  Then for sufficiently large $N$ and small $p$, the exponent governing the probability in Theorem \ref{thm:pFixed} is of order $N p^{\gamma} \epsilon = \omega\left( N p^{2\gamma} \right) = \omega\left( N^{1-2\gamma} \right)$ and thus increases without bound as desired.  Assume then that $\epsilon = O\left( p^{\gamma} \right)$.  Then the exponent is asymptotically of order $N \epsilon^{2}$, which requires that $\epsilon = \omega\left( 1/\sqrt{N} \right)$ for the bound \eqref{eqn:G0} to hold with probability converging to $1$.

For the case $1/2 < \gamma < 1$, first note that the $p^{1-\gamma}$ term in the denominator of the exponent dominates the $p^{\gamma}$ term for small $p$.  If $\epsilon = \omega\left( p^{1-\gamma} \right)$, then the exponent is again asymptotically of order $N p^{\gamma} \epsilon = \omega(Np)$ and increases to infinity.  If $\epsilon = O\left( p^{1-\gamma} \right)$, then the order of the exponent becomes $N p^{2\gamma-1} \epsilon^{2}$, requiring that $\epsilon$ decrease according to the stated rate.  
\end{IEEEproof}

Using Corollary \ref{cor:pVanishing}, we examine next the limits at low and high SNR/sensing budget as in Section \ref{subsec:pFixed}.  The result below specifies the asymptotic dependence of the two-stage gain on the ratio $r$ as well as the asymptotically optimal division of the sensing budget between the first and second stages.  A proof is given in Appendix \ref{app:pfpVanishing}.

\begin{theorem}\label{thm:pVanishing}
Assume that $Np \to \infty$ and $p \to 0$ as in Corollary \ref{cor:pVanishing}.  For low SNR or sensing budgets, 
the optimal two-stage gain 
satisfies
\[
\lim_{\substack{Np\to\infty\\p\to 0}} G \geq 1 + \frac{(1-\gamma) s r^{2}}{8} + O\left(r^{3}\right)
\]
and the optimal fraction of the sensing budget allocated to the first stage approaches $\lambda^{\ast} = 1/2$.  For high SNR or sensing budgets, 
the optimal two-stage gain increases as 
\[
\lim_{\substack{Np\to\infty\\p\to 0}} G \geq C_{3}(q) e^{s/2} \sqrt{r} + O\left(r^{\max\left\{\frac{1}{2}-\frac{1}{q},-\frac{1}{2}\right\}}\right),
\]
where
\[
C_{3}(q) = \frac{q^{\frac{3q+2}{4}}}{(q+2)^{\frac{q+2}{4}} (q+1)^{\frac{q+1}{2}}},
\]
and the optimal first-stage fraction approaches $\lambda^{\ast} = 1/(q+1)$.
\end{theorem}


Theorem \ref{thm:pVanishing} shows that the gain increases with the SNR/sensing budget according to $\sqrt{r}$, independent of the loss exponent $q$.  Another difference with the fixed-$p$ result in Theorem \ref{thm:pFixedHighSNR} is that the first-stage budget fraction $\lambda^{\ast}$ does not decrease to zero as $r \to \infty$.  This can again be explained by the faster increase of the sensing budget relative to the number of nonzero components.  As a result, arbitrarily high gains can be achieved while using only a fraction of the budget in the second exploitation stage, leaving more resources for detecting nonzero components in the first stage to increase the gains in the second stage.  Optimizing this tradeoff yields the solution $\lambda^{\ast} = 1/(q+1)$ in the theorem.

\section{Numerical validation}
\label{sec:numVal}

The results in Section \ref{subsec:pFixed} for finite sparsity levels $p$ 
are validated through numerical simulations.  The signal dimension $N$ is set to $10000$, the prior signal standard deviation $\sigma$ is normalized to $1$, and the prior signal mean $\mu$ is set to $4$ corresponding to $s = 16$.  

\begin{figure*}[!t]
\centerline{
\subfigure[]{\includegraphics[width=0.33\textwidth]{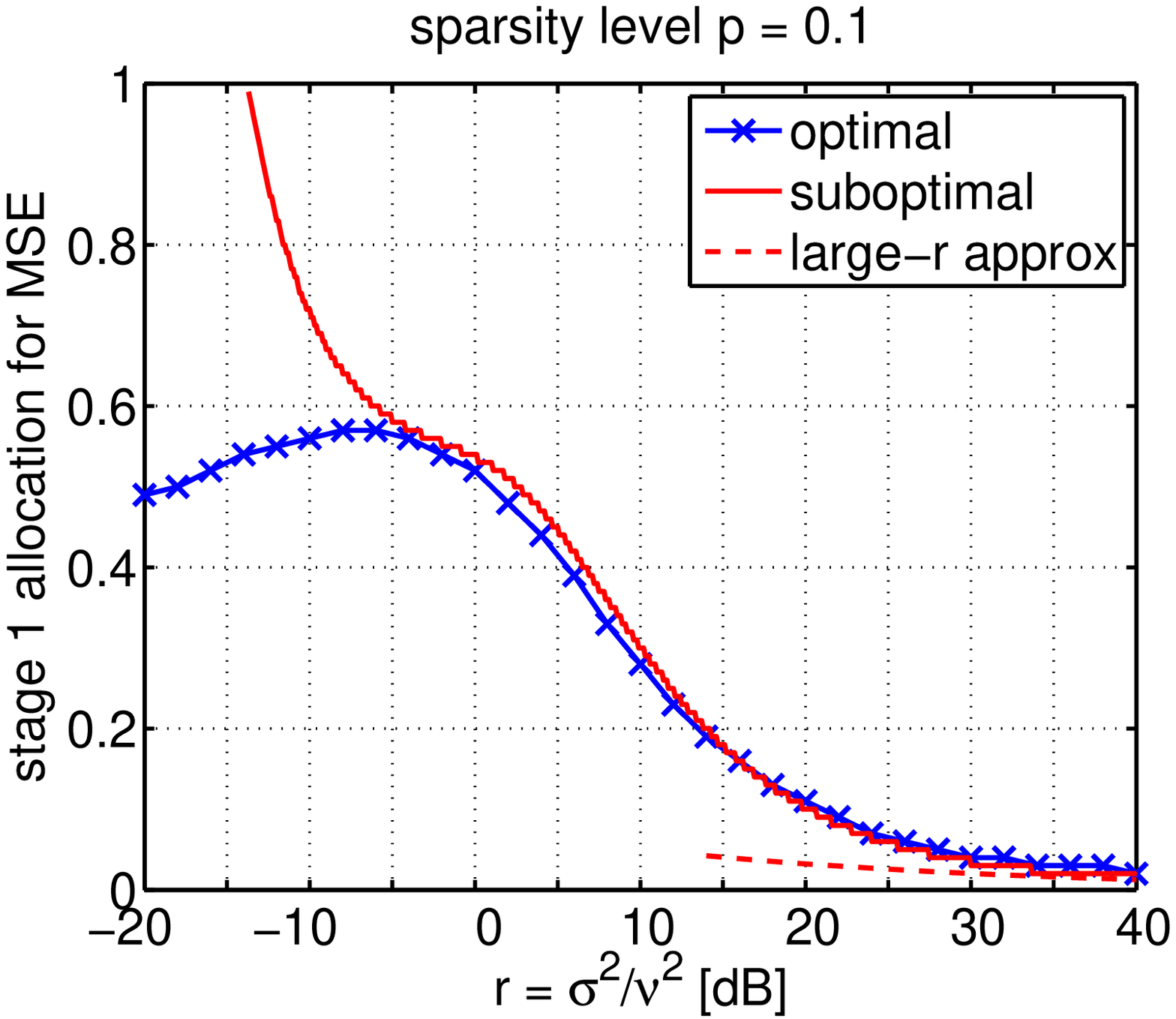}
\label{fig:lambdap1c1}}
\subfigure[]{\includegraphics[width=0.33\textwidth]{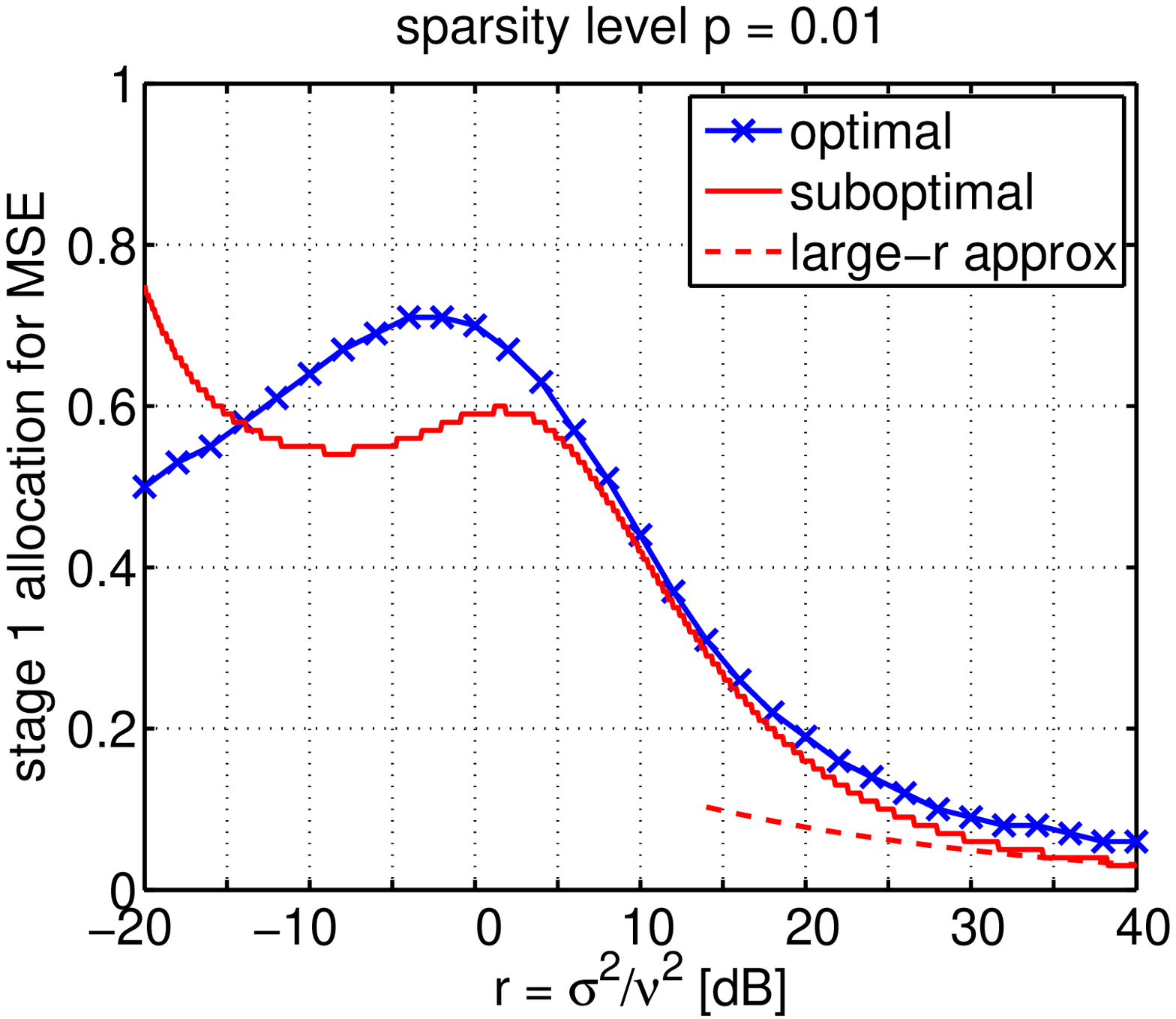}
\label{fig:lambdap2c1}}
\subfigure[]{\includegraphics[width=0.33\textwidth]{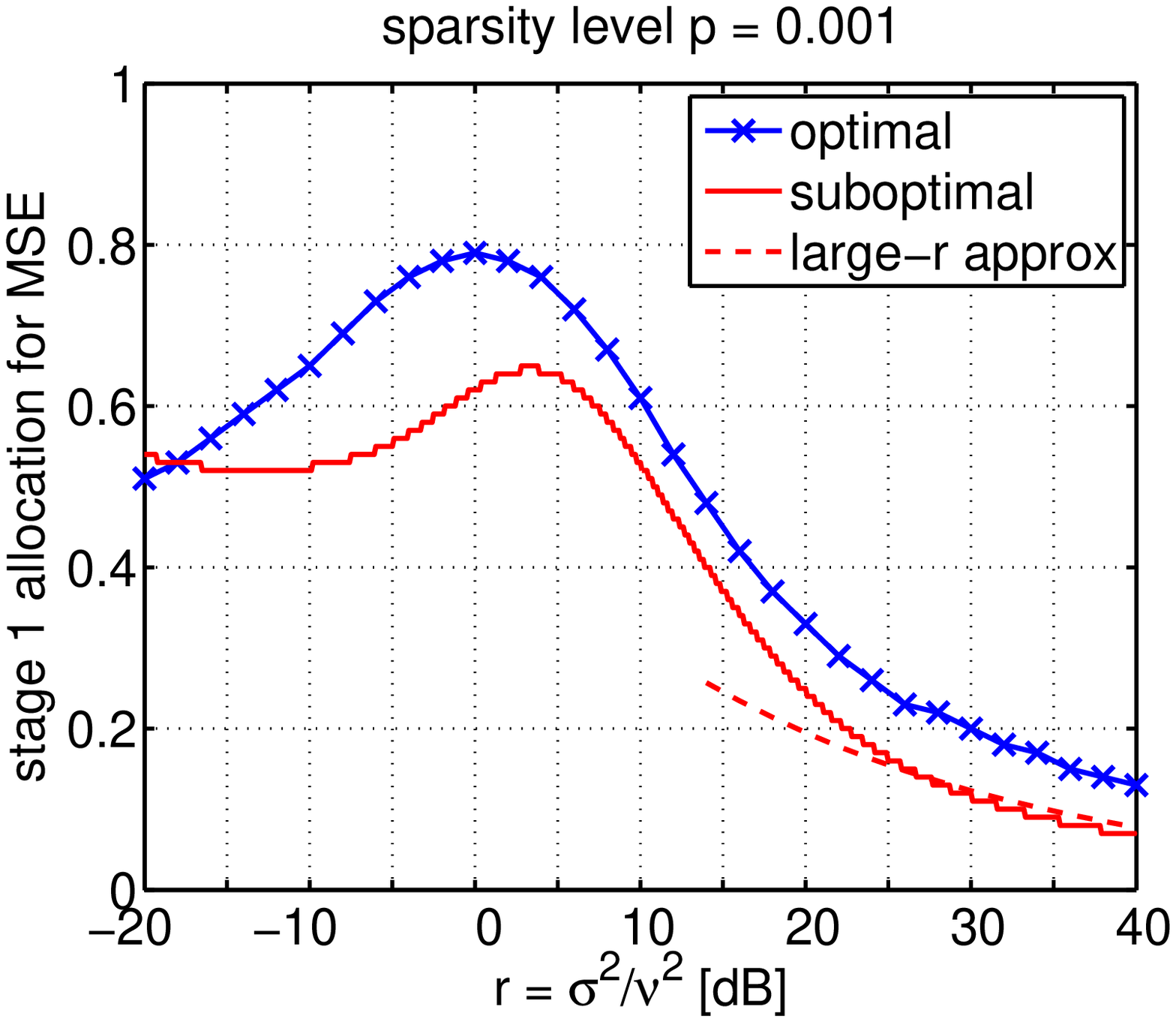}
\label{fig:lambdap3c1}}}
\centerline{
\subfigure[]{\includegraphics[width=0.33\textwidth]{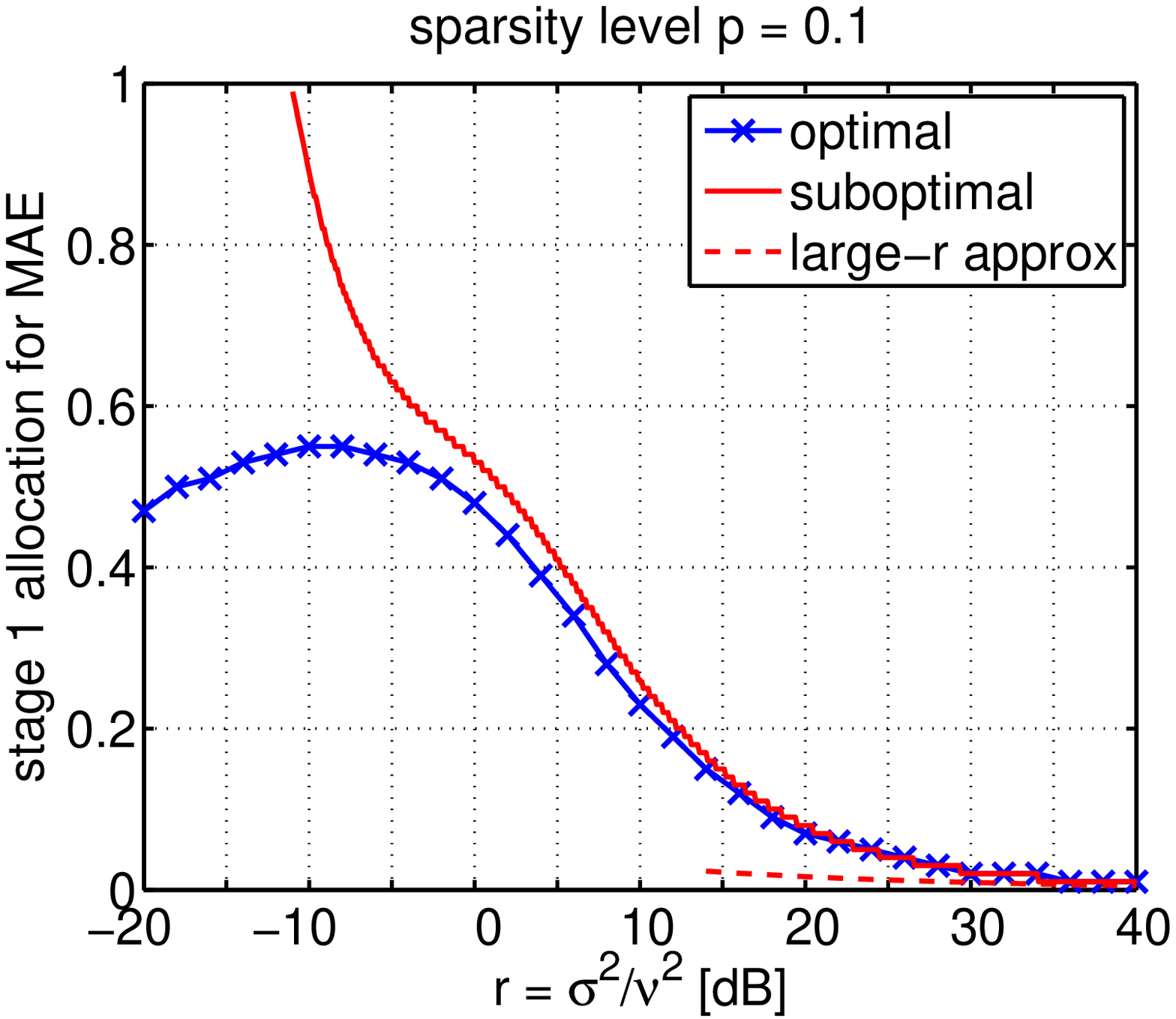}
\label{fig:lambdap1c2}}
\subfigure[]{\includegraphics[width=0.33\textwidth]{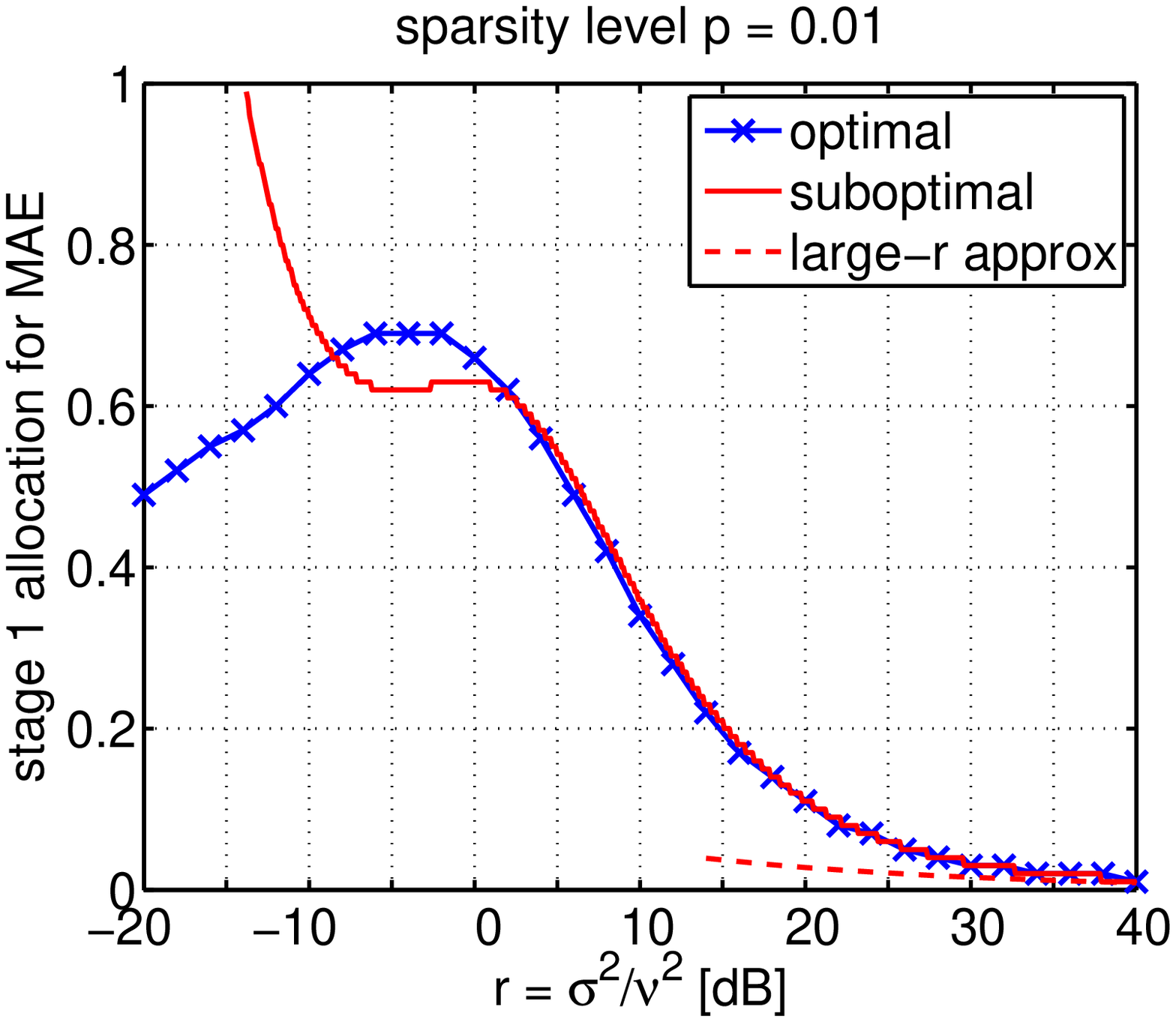}
\label{fig:lambdap2c2}}
\subfigure[]{\includegraphics[width=0.33\textwidth]{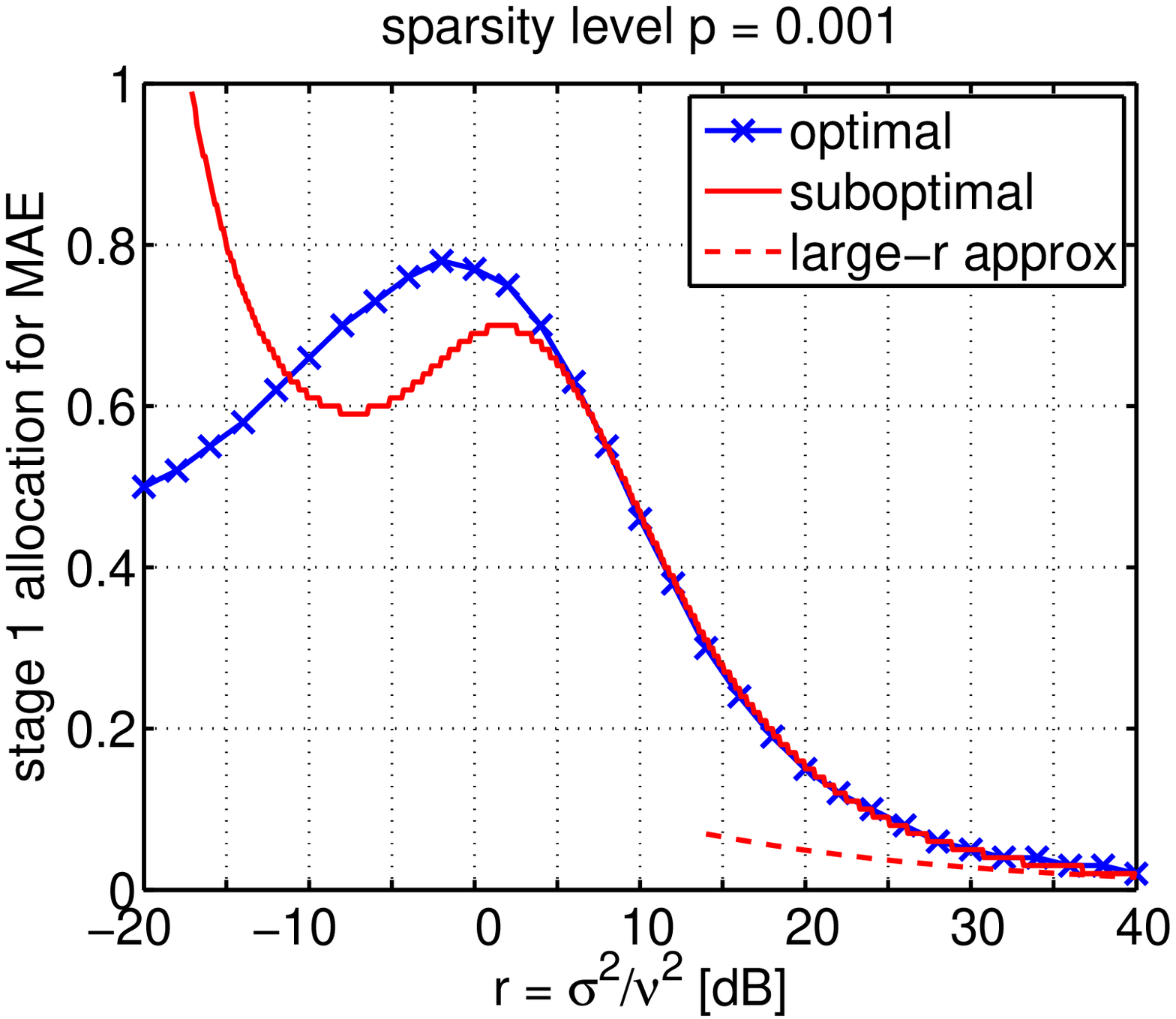}
\label{fig:lambdap3c2}}}
\caption{First-stage budget allocation $\lambda$ as a function of SNR/sensing budget parameter $r = \sigma^{2}/\nu^{2}$ for sparsity levels $p = 0.1, 0.01, 0.001$ and either MSE (top row, $q = 2$ in \eqref{eqn:MQEROI}) or MAE (bottom row, $q = 1$).  The suboptimal values of $\lambda$ that optimize the upper bound \eqref{eqn:J0*UBpInf} (using Propositions \ref{prop:Cp}, \ref{prop:BCp} to bound $C_{p}^{\gamma}$) agree well with the optimal values from \eqref{eqn:J0*} for moderate and large $r$.  The large-$r$ approximation to the suboptimal allocation becomes accurate for $r$ greater than $30$ dB.}
\label{fig:lambda}
\end{figure*}

We discuss first the determination of the fraction $\lambda$ of the sensing budget allocated to the first stage.  Fig.~\ref{fig:lambda} compares the optimal $\lambda$ from \eqref{eqn:J0*} to the suboptimal choice obtained by optimizing the upper bound \eqref{eqn:J0*UBpInf} 
for a range of sparsity levels and SNR/sensing budgets.  We consider both MSE ($q = 2$ in \eqref{eqn:MQEROI}) and mean absolute error (MAE, $q = 1$), using Proposition \ref{prop:BCp} in the former case and the first inequality \eqref{eqn:CpUB1} in Proposition \ref{prop:Cp} in the latter case to bound the Chernoff coefficient $C_{p}^{\gamma}$.  The plots show generally good agreement between the optimal and suboptimal values for moderate and large $r$.  For small $r$, the values deviate because of the decreasing quality of the approximations to $C_{p}^{\gamma}$ in Propositions \ref{prop:Cp} and \ref{prop:BCp}.  In some cases for very small $r$, the best upper bound on $C_{p}^{\gamma}$ is the trivial bound of $1$, which makes the right-hand side of \eqref{eqn:J0*UBpInf} independent of $\lambda$ and hence prevents $\lambda$ from being determined.  These cases appear as gaps in the suboptimal curves in Fig.~\ref{fig:lambda}.  We also show in Fig.~\ref{fig:lambda} the large-$r$ approximations \eqref{eqn:lambda*HighSNR} to the suboptimal first-stage allocations.  The approximations become increasingly accurate for $r$ greater than $30$ dB.


\begin{figure*}[!t]
\centerline{
\subfigure[]{\includegraphics[width=0.33\textwidth]{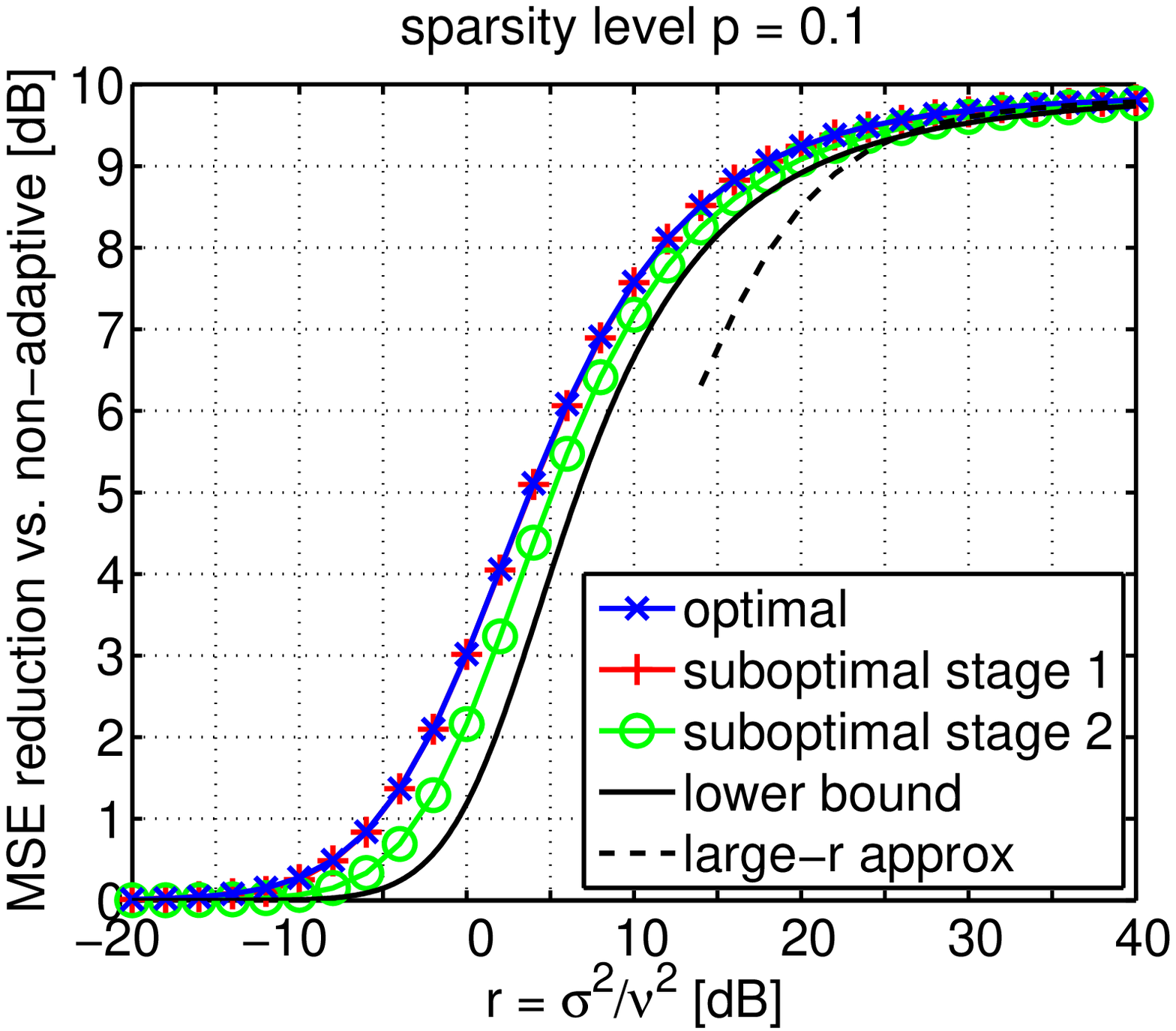}
\label{fig:gainp1c1}}
\subfigure[]{\includegraphics[width=0.33\textwidth]{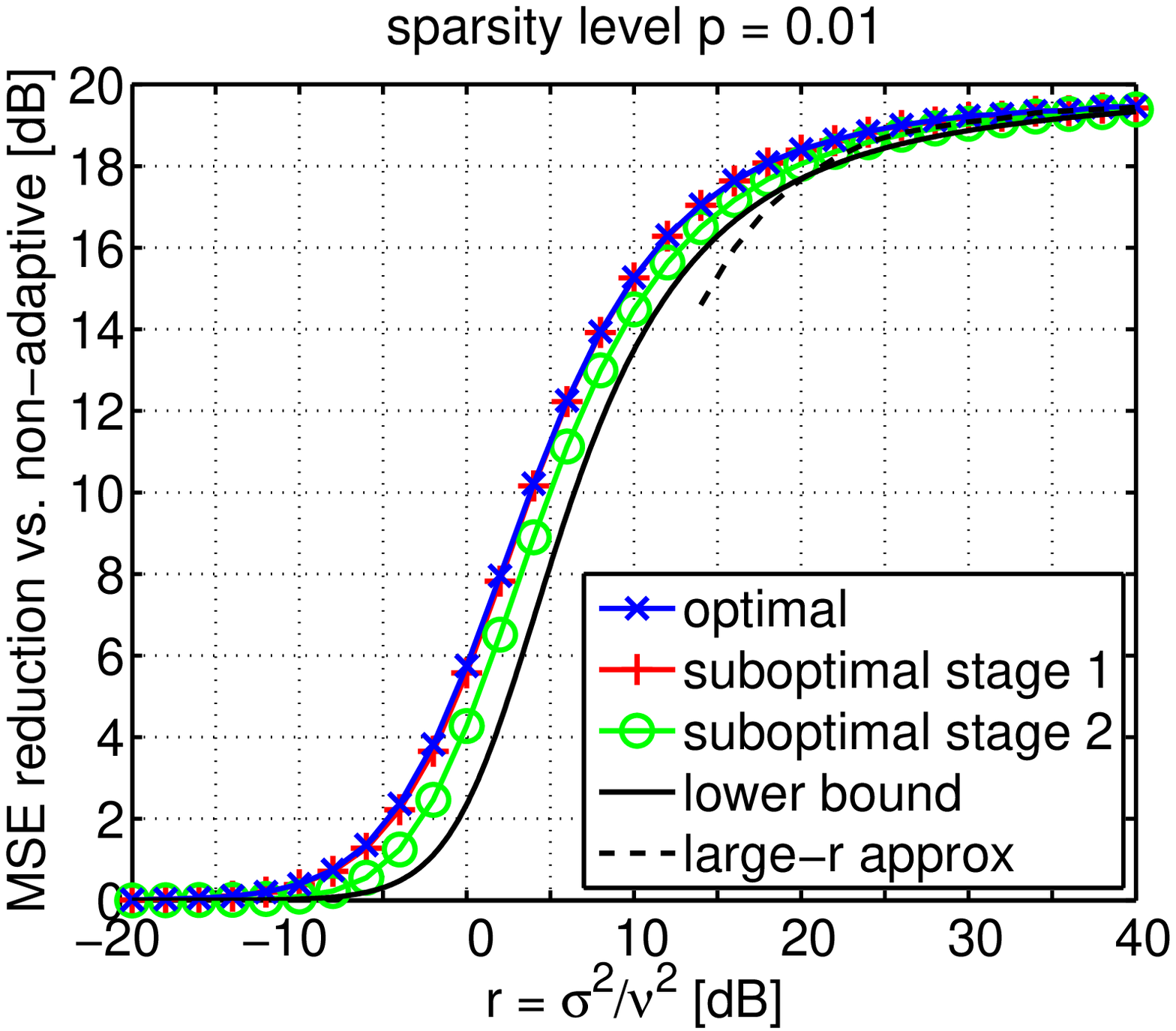}
\label{fig:gainp2c1}}
\subfigure[]{\includegraphics[width=0.33\textwidth]{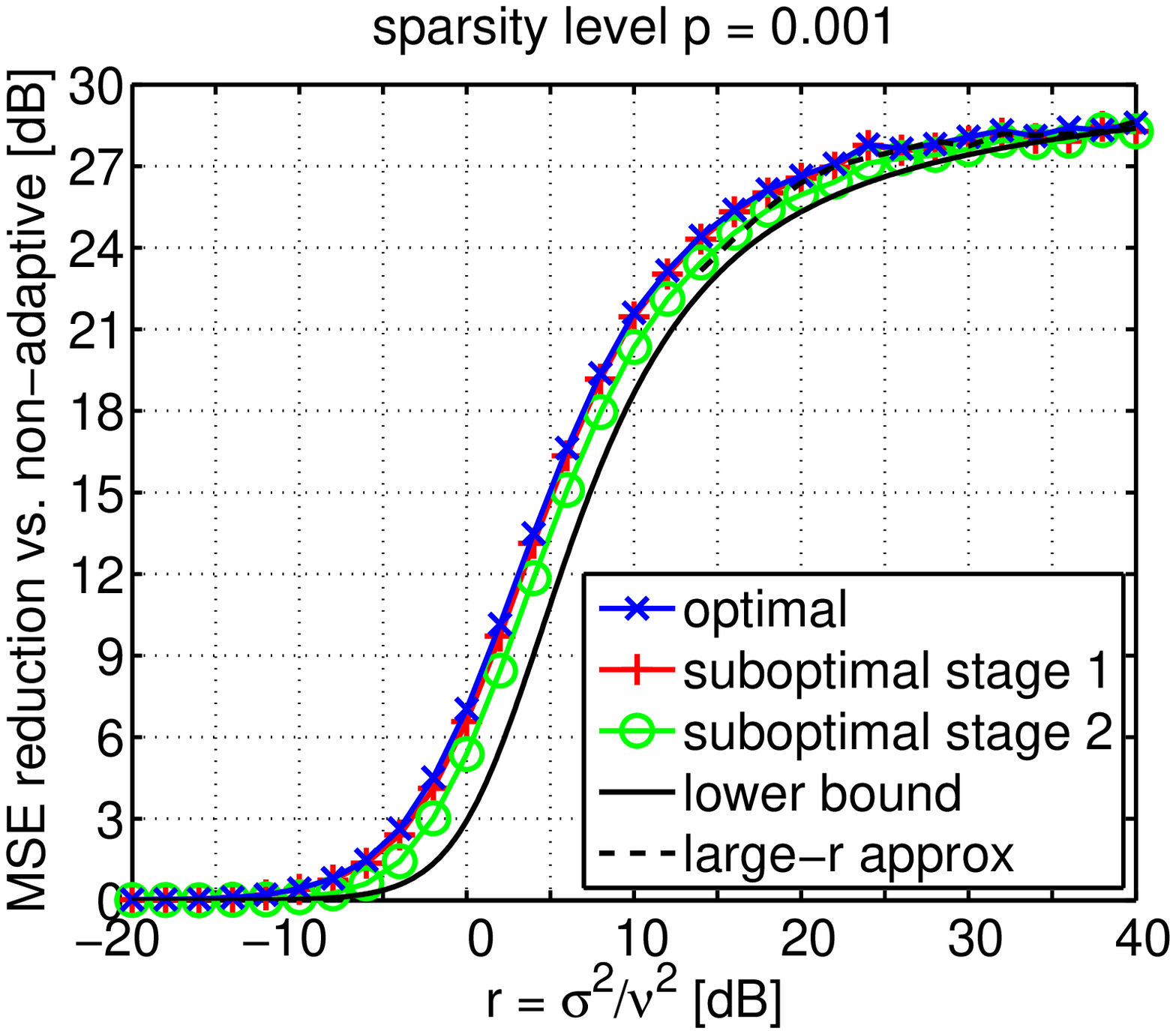}
\label{fig:gainp3c1}}}
\centerline{
\subfigure[]{\includegraphics[width=0.33\textwidth]{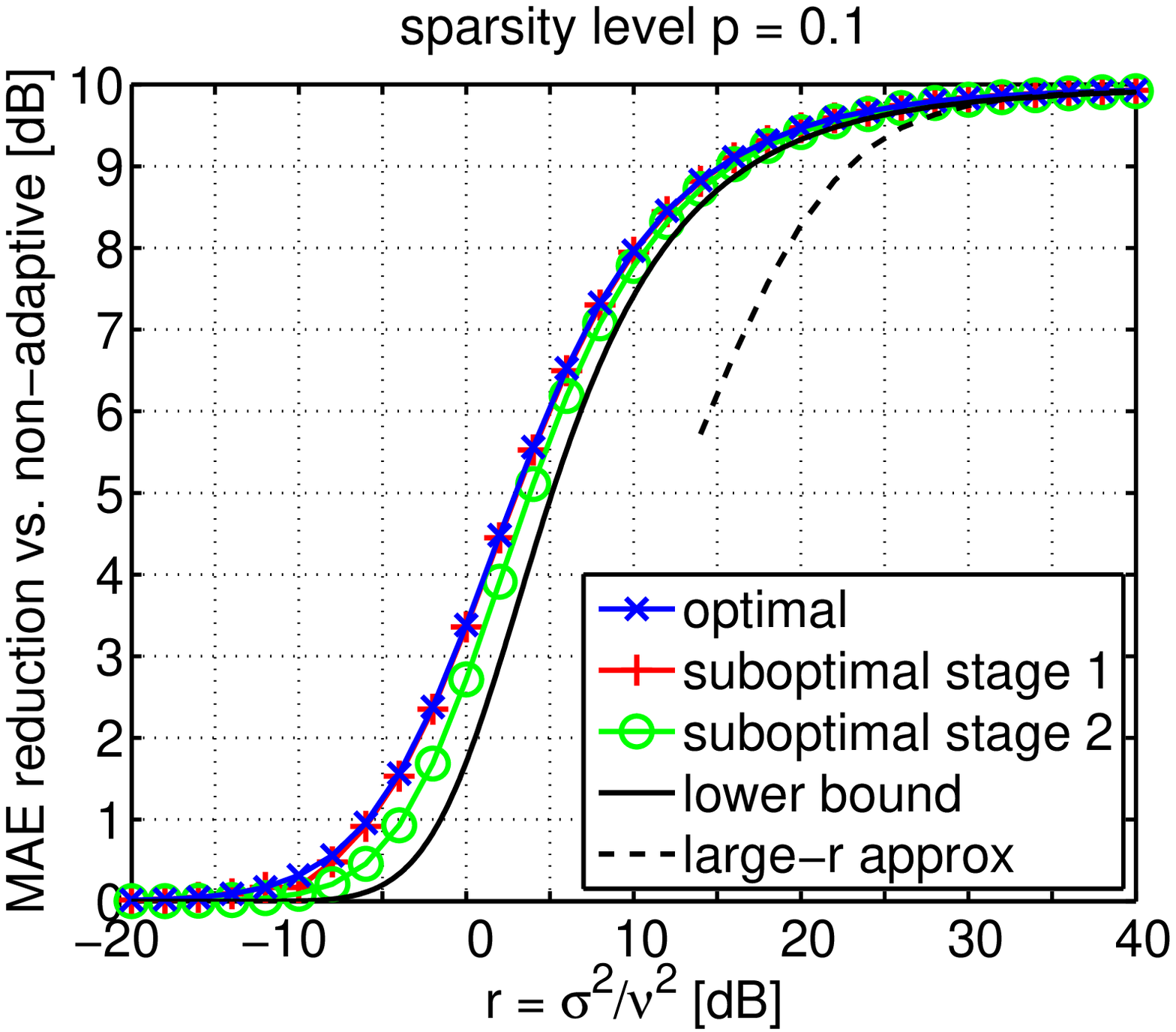}
\label{fig:gainp1c2}}
\subfigure[]{\includegraphics[width=0.33\textwidth]{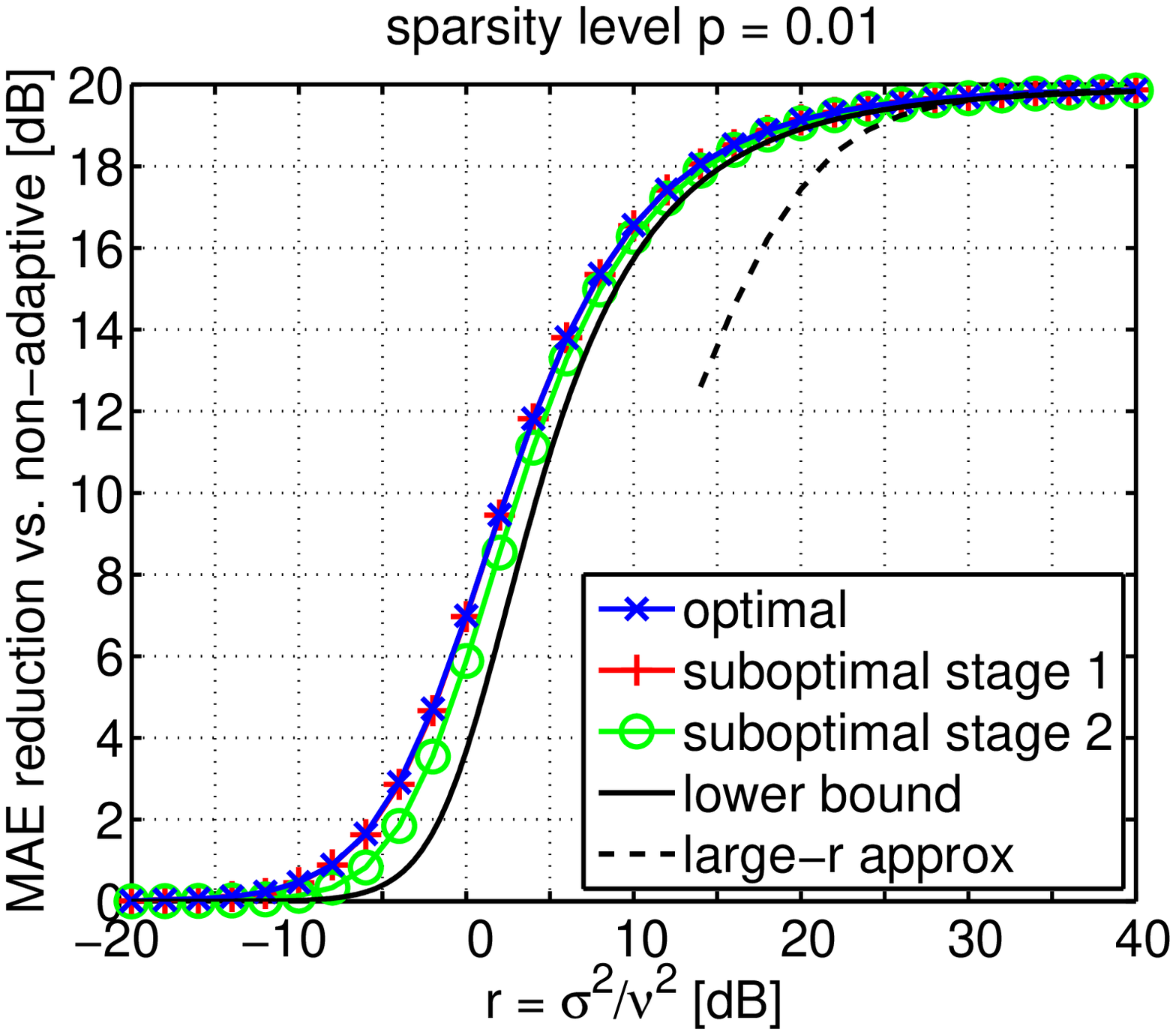}
\label{fig:gainp2c2}}
\subfigure[]{\includegraphics[width=0.33\textwidth]{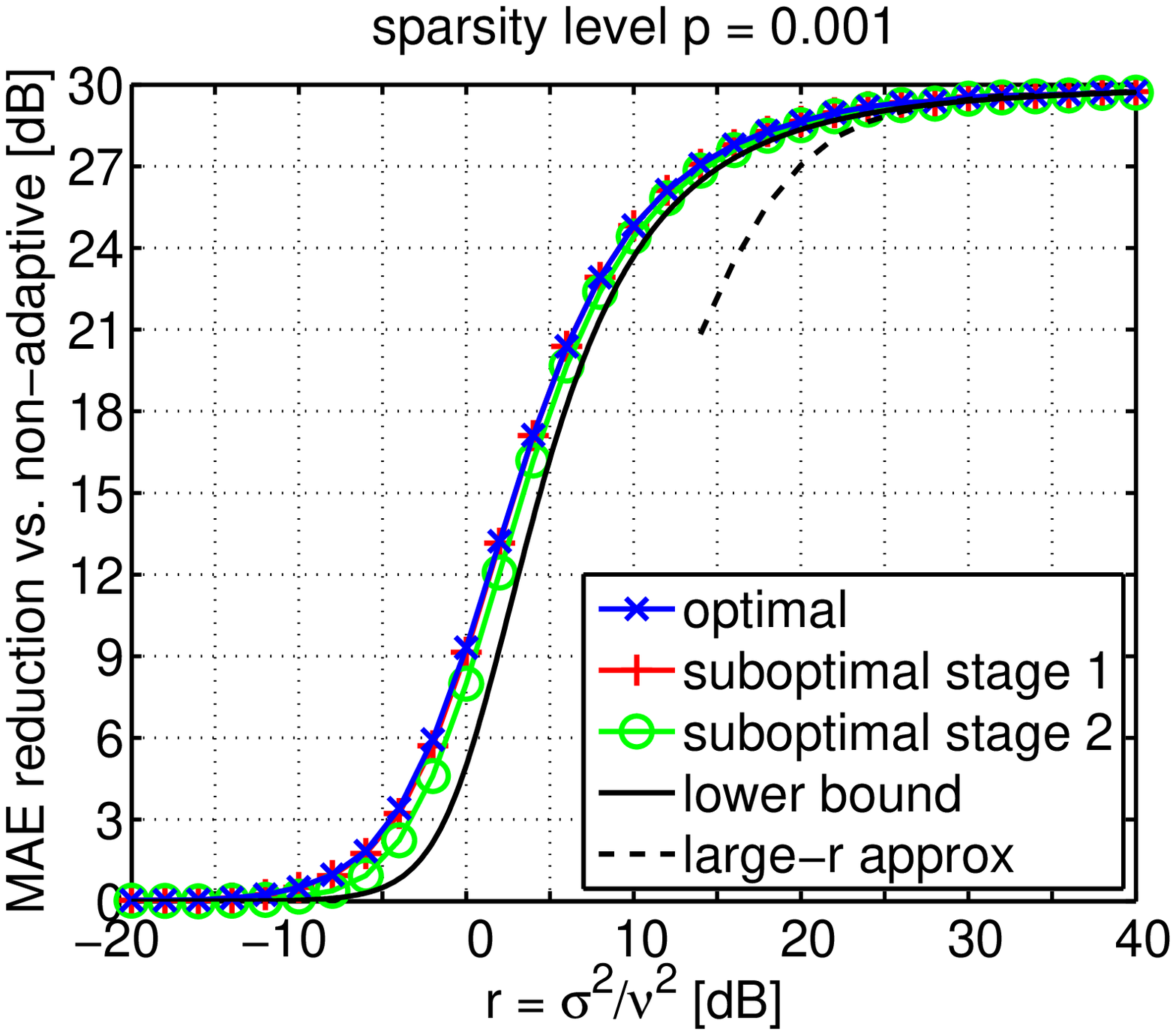}
\label{fig:gainp3c2}}}
\caption{Gain (MSE or MAE reduction) relative to non-adaptive sensing as a function of SNR/sensing budget parameter $r = \sigma^{2}/\nu^{2}$ for sparsity levels $p = 0.1, 0.01, 0.001$.  The lower bound resulting from \eqref{eqn:Gp} and Propositions \ref{prop:Cp}, \ref{prop:BCp} closely follows the gain of the optimal two-stage policy, with both curves converging to $1$ and $(1/p)^{q/2}$ at small and large $r$ respectively.  The gain of the policy with suboptimal first stage is nearly identical to the optimal gain while the gain of the policy with suboptimal second stage lies roughly midway between the optimal gain and the lower bound.  The large-$r$ approximate policy approaches the optimal gain above $r = 25$ dB.} 
\label{fig:gain}
\end{figure*}

In Fig.~\ref{fig:gain}, we compare the adaptation gain, i.e., the reduction in MSE or MAE relative to non-adaptive uniform sensing, of the optimal two-stage policy in Section \ref{subsec:policyOpt} to the lower bound \eqref{eqn:Gp} in Theorem \ref{thm:pFixed}, again combined with either Propositions \ref{prop:Cp} or \ref{prop:BCp} to bound $C_{p}^{\gamma}$.  All empirical means in Fig.~\ref{fig:gain} are computed from $10000$ simulations. 
It is seen that the lower bound, which is an asymptotic result, approximates the finite-dimensional optimal gain very well.  At small and large $r$, both curves approach gains of $1$ and $(1/p)^{q/2}$ respectively as predicted.  The maximum deviations of the two curves occur at intermediate $r$ near unity and are $1.9$, $3.6$, $4.6$ dB for $p = 0.1$, $0.01$, $0.001$ and MSE, and $1.7$, $3.3$, $4.3$ dB for the same $p$ values and MAE.  Fig.~\ref{fig:gain} also plots the gains of 
three suboptimal policies, the first using the suboptimal first-stage allocation in Fig.~\ref{fig:lambda} and the optimal form \eqref{eqn:pi}--\eqref{eqn:b} for the second stage, the second using the suboptimal form \eqref{eqn:lambdaSubopt} for the second stage, and the third using the large-$r$ approximation \eqref{eqn:lambda*HighSNR} to the first-stage and optimal second stage.  The first of these policies 
achieves gains that are nearly identical to the optimal gains, while the policy with suboptimal second stage shows that roughly one third to one half of the gap between the optimal gain and the lower bound \eqref{eqn:Gp} is due to the simpler functional form assumed in \eqref{eqn:lambdaSubopt}.  The large-$r$ approximate policy approaches the optimal gain above $r = 25$ dB, suggesting that the lowest-order approximation \eqref{eqn:lambda*HighSNR} is sufficiently accurate in this regime.  The approximation is better for MSE and improves as $p$ decreases. 


\section{Conclusion and future work}
\label{sec:concl}

This paper has provided analytical guarantees on the performance of two-stage adaptive sensing for estimating the nonzero amplitudes in a sparse signal.  The results apply to the mean $q$th-power estimation error metric for any positive $q$.  The improvement compared to non-adaptive sensing is shown to be related to the detectability of nonzero components as measured by Chernoff coefficients, thus quantifying the dependence on the sparsity level, the SNR, and the sensing resource budget.  In the case of fixed sparsity level and increasing SNR or sensing budget, we have derived the rate of convergence to oracle performance as well as the rate at which the fraction of the budget devoted to signal support identification decays to zero.  In the case of a vanishing fraction of nonzero components, the rate of increase of the adaptation gain with SNR/sensing budget is quantified.  Numerical simulations have demonstrated that the bounds yield good approximations to the optimal two-stage policy and can thus be used to quantitatively predict performance. 

Future work will consider the extension of the techniques in this paper to analogous policies for adaptive spectrum sensing \cite{adapEstICASSP2013}, to other tasks such as signal support recovery, and to more refined analysis of policies with more than two stages.  It is anticipated that the extension to multiple stages will require the ability to handle non-uniform belief states $\mbs(t)$, e.g.~non-uniform signal presence probabilities $p_{i}(t)$, which arise once the first stage of observations has been incorporated.  The current work by contrast assumes a uniform signal prior and first-stage allocation $\lambda$, which makes the first-stage observations $y_{i}(1)$ identically distributed and hence simpler to analyze.  Lastly, it would be desirable to determine fundamental limits on adaptive sensing for the problem considered in this work, i.e., upper bounds on performance along the lines of \cite{ariascastro2013,castro2013} that are better than an oracle bound.

\appendices
\numberwithin{equation}{section}


\section{Tail bound}
\label{app:tailBounds}

In this appendix, we derive a tail bound on a sample average that appears in the proof of Theorem \ref{thm:pFixed}.

\begin{lemma}\label{lem:psqrt}
For $\gamma \in (0,1)$ and any $\epsilon > 0$, 
\[
\frac{1}{N} \sum_{i=1}^{N} p_{i}^{\gamma}(1) \leq (1+\epsilon) p^{\gamma} C_{p}^{\gamma}
\]
with probability at least
\[
\begin{cases}
1 - \exp \left( -\dfrac{\left(C_{p}^{\gamma}\right)^{2} N p^{\gamma} \epsilon^{2}}{2 \left(p^{\gamma} \left(C_{p}^{2\gamma} - \left(C_{p}^{\gamma}\right)^{2}\right) + \epsilon C_{p}^{\gamma} /3\right)} \right), & 0 < \gamma \leq \frac{1}{2},\\
1 - \exp \left( -\dfrac{\left(C_{p}^{\gamma}\right)^{2} N p^{\gamma} \epsilon^{2}}{2 \left(p^{1-\gamma} - p^{\gamma} \left(C_{p}^{\gamma}\right)^{2} + \epsilon C_{p}^{\gamma} /3\right)} \right), & \frac{1}{2} < \gamma < 1.
\end{cases}
\]
\end{lemma}
\begin{IEEEproof}
From \eqref{eqn:pEvol} and \eqref{eqn:fp} we have 
%
\begin{equation}\label{eqn:p1}
p_{i}(1) = \frac{p f_{1}(y_{i}(1))}{p f_{1}(y_{i}(1)) + (1-p) f_{0}(y_{i}(1))} = \frac{p f_{1}(y_{i}(1))}{f_{p}(y_{i}(1))},
\end{equation}
and since $\{ y_{i}(1) \}_{i=1}^{N}$ are i.i.d., so too are $\left\{ p_{i}^{\gamma}(1) \right\}_{i=1}^{N}$.  
The expected value of $p_{i}^{\gamma}(1)$ is given by 
%
\begin{equation}\label{eqn:pgammaMean}
\E \left[ p_{i}^{\gamma}(1) \right] = \int_{-\infty}^{\infty} p^{\gamma} \left(\frac{f_{1}(y)}{f_{p}(y)}\right)^{\gamma} f_{p}(y) dy = p^{\gamma} C_{p}^{\gamma}
\end{equation}
using \eqref{eqn:Cp}.  For $0 < \gamma \leq 1/2$, the second moment of $p_{i}^{\gamma}(1)$ is likewise given by $p^{2\gamma} C_{p}^{2\gamma}$, while for $1/2 < \gamma < 1$, we use the fact that $p_{i}(1) \leq 1$ as a probability to bound the second moment: 
\[
\E \left[ p_{i}^{2\gamma}(1) \right] \leq \E \left[ p_{i}(1) \right] 
= p.
\]
Hence 
\begin{equation}\label{eqn:pgammaVar}
\var\left( p_{i}^{\gamma}(1) \right) 
\begin{cases}
= p^{2\gamma} \left( C_{p}^{2\gamma} - \left(C_{p}^{\gamma}\right)^{2} \right), & 0 < \gamma \leq \frac{1}{2},\\
\leq p - p^{2\gamma} \left(C_{p}^{\gamma}\right)^{2}, & \frac{1}{2} < \gamma < 1.
\end{cases}
\end{equation}
%
We may now invoke Bernstein's inequality 
for random variables bounded by $M$ \cite{vandervaart2000}: 
\[
\mbbP\left( \frac{1}{N} \sum_{i=1}^{N} p_{i}^{\gamma}(1) > \E \left[ p_{i}^{\gamma}(1) \right] + \delta \right) \leq \exp\left( -\frac{N\delta^{2}}{2 \left(\var\left( p_{i}^{\gamma}(1) \right) + M \delta/3 \right)} \right),
\]
with $M = 1$.  
Substituting \eqref{eqn:pgammaMean} and \eqref{eqn:pgammaVar} and letting $\delta = \epsilon p^{\gamma} C_{p}^{\gamma}$ establishes the result.
\end{IEEEproof}

\section{Bounds on Chernoff coefficients}
\label{app:pfCp}

We prove Proposition \ref{prop:Cp}, which provides upper bounds on the Chernoff coefficient $C_{p}^{\gamma}$ \eqref{eqn:Cp} in terms of $C_{0}^{\gamma}$ \eqref{eqn:C0}, and Proposition \ref{prop:BCp}, which gives a tighter upper bound for the special case of the Bhattacharyya coefficient $C_{p}^{1/2}$.  

\begin{IEEEproof}[Proof of Proposition \ref{prop:Cp}]
We use the fact that the function $z^{1-\gamma}$ is 
concave in $z \geq 0$ for $\gamma \in (0,1)$.  Hence its value at $z = z_{1} + z_{2}$ with $z_{1} > z_{2} \geq 0$ is bounded above by its linear approximation at $z = z_{1}$:
\[
\left(z_{1} + z_{2}\right)^{1-\gamma} \leq z_{1}^{1-\gamma} + (1-\gamma) z_{1}^{-\gamma} z_{2}.
\]
%
For $y \in \mcY_{0}$, we let $z_{1} = (1-p) f_{0}(y)$ and $z_{2} = p f_{1}(y)$ to obtain 
\begin{align}
\left(p f_{1}(y) + (1-p) f_{0}(y)\right)^{1-\gamma} &\leq (1-p)^{1-\gamma} f_{0}^{1-\gamma}(y) + (1-\gamma) \frac{p f_{1}(y)}{((1-p) f_{0}(y))^{\gamma}}\nonumber\\
&\leq (1-p)^{1-\gamma} f_{0}^{1-\gamma}(y) + (1-\gamma) p^{1-\gamma} f_{1}^{1-\gamma}(y),\label{eqn:CpY0}
\end{align}
where the second inequality follows from the definition of $\mcY_{0}$.  For $y \in \mcY_{1}$ we reverse the roles of $(1-p) f_{0}(y)$ and $p f_{1}(y)$, yielding 
\begin{equation}\label{eqn:CpY1}
\left(p f_{1}(y) + (1-p) f_{0}(y)\right)^{1-\gamma} \leq p^{1-\gamma} f_{1}^{1-\gamma}(y) + (1-\gamma) \frac{(1-p) f_{0}(y)}{(p f_{1}(y))^{\gamma}}.
\end{equation}
Applying \eqref{eqn:CpY0} and \eqref{eqn:CpY1} pointwise to the integrand in \eqref{eqn:Cp} results in 
\begin{align}
C_{p}^{\gamma} &\leq (1-p)^{1-\gamma} \int_{\mcY_{0}} f_{1}^{\gamma}(y) f_{0}^{1-\gamma}(y) dy + (1-\gamma) p^{1-\gamma} \int_{\mcY_{0}} f_{1}(y) dy\nonumber\\ 
&\qquad + p^{1-\gamma} \int_{\mcY_{1}} f_{1}(y) dy + (1-\gamma) \frac{1-p}{p^{\gamma}} \int_{\mcY_{1}} f_{0}(y) dy\label{eqn:CpUB0}\\
&= p^{1-\gamma} \left(1-\gamma + \gamma \mbbP_{1}(\mcY_{1})\right) + (1-p)^{1-\gamma} \left( \int_{\mcY_{0}} f_{1}^{\gamma}(y) f_{0}^{1-\gamma}(y) dy + (1-\gamma) \left(\frac{1-p}{p}\right)^{\gamma} \mbbP_{0}(\mcY_{1}) \right).\label{eqn:CpUB}
\end{align}
Using \eqref{eqn:f0f1}, \eqref{eqn:C0} and the definition of $r$ and after some straightforward algebra, we find that 
\begin{equation}\label{eqn:C0Partial}
\int_{\mcY_{0}} f_{1}^{\gamma}(y) f_{0}^{1-\gamma}(y) dy = C_{0}^{\gamma} \int_{\mcY_{0}} f_{01}(y) dy = \mbbP_{01}(\mcY_{0}) C_{0}^{\gamma},
\end{equation}
where $f_{01}$ is a Gaussian PDF with parameters given in the statement of the proposition.  Combining \eqref{eqn:CpUB}, \eqref{eqn:C0Partial}, and the trivial bound $C_{p}^{\gamma} \leq 1$, we arrive at the first bound \eqref{eqn:CpUB1} in the proposition statement.

To derive the second, weaker bound \eqref{eqn:CpUB2} from the first bound, we use the definition of $\mcY_{1}$ to relax \eqref{eqn:CpY1} as follows:
\begin{align}
\left(p f_{1}(y) + (1-p) f_{0}(y)\right)^{1-\gamma} &\leq p^{1-\gamma} f_{1}^{1-\gamma}(y) + (1-\gamma) (1-p)^{1-\gamma} f_{0}^{1-\gamma}(y)\nonumber\\
&\leq p^{1-\gamma} f_{1}^{1-\gamma}(y) + (1-p)^{1-\gamma} f_{0}^{1-\gamma}(y).\label{eqn:CpY12}
\end{align}
Note that \eqref{eqn:CpY12} is also a relaxation of \eqref{eqn:CpY0}.  Substituting \eqref{eqn:CpY12} into \eqref{eqn:Cp} and combining with the trivial bound $C_{p}^{\gamma} \leq 1$ yields \eqref{eqn:CpUB2}. 

As $r\lambda$ increases, the densities $f_{0}(y)$ and $f_{1}(y)$ become more concentrated and dominate the other density within their respective regions $\mcY_{0}$ and $\mcY_{1}$.  As a consequence, the linear approximations in \eqref{eqn:CpY0}, \eqref{eqn:CpY1} to the function $z^{1-\gamma}$ become more accurate, with the first terms on the right-hand sides of \eqref{eqn:CpY0}, \eqref{eqn:CpY1} becoming dominant.  Correspondingly, the first and third terms on the right-hand side of \eqref{eqn:CpUB0} converge to $(1-p)^{1-\gamma} C_{0}^{\gamma}$ and $p^{1-\gamma}$ respectively, while the second and fourth terms become negligible.  This shows that the second inequality \eqref{eqn:CpUB2} becomes tight as $r\lambda$ increases, and by extension the first inequality \eqref{eqn:CpUB1} as well.  In the limit $r\lambda \to \infty$, both $C_{p}^{\gamma}$ and the upper bounds converge to $p^{1-\gamma}$. 

It remains to specify the probabilities $\mbbP_{1}(\mcY_{1})$, $\mbbP_{01}(\mcY_{0})$, and $\mbbP_{0}(\mcY_{1})$ in terms of the standard Gaussian CDF $\Phi$.  First we determine the boundaries between the regions $\mcY_{0}$ and $\mcY_{1}$ by setting $(1-p) f_{0}(y) = p f_{1}(y)$ and using \eqref{eqn:f0f1}:
\begin{align*}
\frac{1-p}{\sqrt{2\pi \nu^{2}/\lambda}} \exp\left( -\frac{y^{2}}{2\nu^{2}/\lambda} \right) &= \frac{p}{\sqrt{2\pi (\sigma^{2} + \nu^{2}/\lambda)}} \exp\left( -\frac{(y-\mu)^{2}}{2(\sigma^{2} + \nu^{2}/\lambda)} \right),\\
\frac{1-p}{p} \sqrt{\frac{\sigma^{2} + \nu^{2}/\lambda}{\nu^{2}/\lambda}} &= \exp\left( \frac{\sigma^{2}}{2(\nu^{2}/\lambda) (\sigma^{2} + \nu^{2}/\lambda)} y^{2} + \frac{\mu}{\sigma^{2} + \nu^{2}/\lambda} y - \frac{\mu^{2}}{2(\sigma^{2} + \nu^{2}/\lambda)} \right),\\
\frac{1-p}{p} \sqrt{1 + r\lambda} &= \exp\left( \frac{sr^{2}\lambda^{2}}{2(1+r\lambda)} \left(\frac{y}{\mu}\right)^{2} + \frac{sr\lambda}{1+r\lambda} \left(\frac{y}{\mu}\right) - \frac{sr\lambda}{2(1+r\lambda)} \right),\\
\frac{1+r\lambda}{s} (2\eta + \log(1 + r\lambda)) &= \left( \frac{r\lambda y}{\mu} \right)^{2} + 2 \left( \frac{r\lambda y}{\mu} \right) - r\lambda,\\
\frac{r\lambda y^{\pm}}{\mu} &= -1 \pm \sqrt{1 + r\lambda + \frac{1+r\lambda}{s} (2\eta + \log(1+r\lambda))},\\
y^{\pm} &= \frac{\mu}{r\lambda} \left( -1 \pm \sqrt{(1+r\lambda)\left(1 + \frac{1}{s} (2\eta + \log(1+r\lambda))\right)} \right).
\end{align*}
In the third line above, we have substituted $r = \sigma^{2}/\nu^{2}$ and $s = \mu^{2}/\sigma^{2}$ as defined in Section \ref{sec:results}, while in the fourth line we have taken the logarithm of both sides and defined $\eta = \log\frac{1-p}{p}$.  The fifth line results from the quadratic formula.  The region $\mcY_{0}$ is the interval $[y^{-}, y^{+}]$ while the region $\mcY_{1}$ is given by $(-\infty, y^{-}) \cup (y^{+}, \infty)$.  

We now standardize $y^{\pm}$ by subtracting the means and dividing by the standard deviations of the Gaussian distributions $f_{1}$, $f_{01}$, and $f_{0}$.  From the relation
\[
\frac{\mu}{\sqrt{\sigma^{2} + \nu^{2}/\lambda}} = \sqrt{\frac{\mu^{2} \lambda / \nu^{2}}{1 + r\lambda}} = \sqrt{\frac{sr\lambda}{1+r\lambda}},
\]
we have 
\begin{align*}
\frac{y^{\pm} - \mu}{\sqrt{\sigma^{2} + \nu^{2}/\lambda}} &= \frac{1}{r\lambda} \sqrt{\frac{sr\lambda}{1+r\lambda}} \left( -(1+r\lambda) \pm \sqrt{(1+r\lambda)\left(1 + \frac{1}{s} (2\eta + \log(1+r\lambda)) \right)} \right)\\
&= \sqrt{\frac{s}{r\lambda}} \left( -\sqrt{1+r\lambda} \pm \sqrt{1 + \frac{1}{s} (2\eta + \log(1+r\lambda)) } \right)\\
&= \mp z_1^{\pm},
\end{align*}
referring to \eqref{eqn:z1}.  Equation \eqref{eqn:P1} then follows from \eqref{eqn:f1}, \eqref{eqn:z1}, and the symmetry of $\Phi$.  Similarly from 
\[
\frac{\mu}{\sqrt{\frac{1+r\lambda}{1+(1-\gamma)r\lambda}\frac{\nu^{2}}{\lambda}}} = \sqrt{\frac{sr\lambda(1+(1-\gamma)r\lambda)}{1+r\lambda}},
\]
we obtain 
\begin{align*}
\frac{y^{\pm} - \frac{\gamma\mu}{1+(1-\gamma)r\lambda}}{\sqrt{\frac{1+r\lambda}{1+(1-\gamma)r\lambda}\frac{\nu^{2}}{\lambda}}} &= \sqrt{\frac{sr\lambda(1+(1-\gamma)r\lambda)}{1+r\lambda}} \left( -\frac{1}{r\lambda} - \frac{\gamma}{1+(1-\gamma)r\lambda} \right.\\
&\qquad\qquad \left. \pm \frac{1}{r\lambda} \sqrt{(1+r\lambda)\left(1 + \frac{1}{s} (2\eta + \log(1+r\lambda)) \right)} \right)\\
&= \sqrt{\frac{sr\lambda(1+(1-\gamma)r\lambda)}{1+r\lambda}} \left( -\frac{1+r\lambda}{r\lambda(1+(1-\gamma)r\lambda)} \right.\\
&\qquad\qquad \left. \pm \frac{1}{r\lambda} \sqrt{(1+r\lambda)\left(1 + \frac{1}{s} (2\eta + \log(1+r\lambda)) \right)} \right)\\
&= \sqrt{\frac{s (1+(1-\gamma)r\lambda)}{r\lambda}} \left( -\frac{\sqrt{1+r\lambda}}{1+(1-\gamma)r\lambda} \pm \sqrt{1 + \frac{1}{s} (2\eta + \log(1+r\lambda))} \right)\\
&= z_{01}^{\pm} 
\end{align*}
from \eqref{eqn:z01}.  Hence $\mbbP_{01}(\mcY_{0})$ is given by \eqref{eqn:P01}.  Lastly,  
\begin{align*}
\frac{y^{\pm}}{\sqrt{\frac{\nu^{2}}{\lambda}}} &= \frac{\mu}{r\lambda \sqrt{\frac{\nu^{2}}{\lambda}}} \left( -1 \pm \sqrt{(1+r\lambda)\left(1 + \frac{1}{s} (2\eta + \log(1+r\lambda))\right)} \right)\\
&= \mp z_{0}^{\pm}
\end{align*}
from \eqref{eqn:z0}.  Combining this with \eqref{eqn:f0} yields \eqref{eqn:P0}. 
\end{IEEEproof}


\begin{IEEEproof}[Proof of Proposition \ref{prop:BCp}]
We use the following bound on the square root appearing in the definition of the Bhattacharyya coefficient $C_{p}^{1/2}$ \eqref{eqn:Cp}. 

\begin{lemma}\label{lem:sqrt}
For $f_{0}, f_{1} \geq 0$ and $p \in [0, 1]$,
\begin{subequations}
\begin{align}
\sqrt{pf_{1} + (1-p)f_{0}} &\leq \sqrt{pf_{1}} + \sqrt{(1-p)f_{0}} - \frac{\sqrt{p(1-p)f_{0}f_{1}}}{\sqrt{pf_{1}} + \sqrt{(1-p)f_{0}}}\label{eqn:sqrt1}\\
&\leq \sqrt{pf_{1}} + \sqrt{(1-p)f_{0}} - \frac{\sqrt{p(1-p)}}{\sqrt{p} + \sqrt{1-p}} \sqrt{\min\{f_{0}, f_{1}\}}.\label{eqn:sqrt2}
\end{align}
\end{subequations}
\end{lemma}
\begin{IEEEproof}
We start from \eqref{eqn:CpY12}, dropping the dependence on $y$ and with $\gamma = 1/2$, and square both sides to yield 
%
\[
p f_{1} + (1-p) f_{0} \leq \left( \sqrt{pf_{1}} + \sqrt{(1-p) f_{0}} \right)^{2}.
\]
%
Instead of taking the square root directly, we exploit the concavity of the square root function as in the proof of Proposition \ref{prop:Cp}, specifically the property that any tangent linear approximation is an overestimator of the function.  For $z_{1}, z_{2} \geq 0$, 
\[
\sqrt{z_{2}} \leq \sqrt{z_{1}} + \frac{1}{2\sqrt{z_{1}}} (z_{2}-z_{1}),
\]
where $1/(2\sqrt{z_{1}})$ is the slope of the tangent at $z_{1}$.  Letting $z_{1} = \left( \sqrt{pf_{1}} + \sqrt{(1-p)f_{0}} \right)^{2}$ and $z_{2} = pf_{1} + (1-p)f_{0}$, we obtain \eqref{eqn:sqrt1}. 
Note that for general $\gamma \neq 1/2$, an analogous argument would result in $z_{1} = \left( (pf_{1})^{1-\gamma} + ((1-p)f_{0})^{1-\gamma} \right)^{\frac{1}{1-\gamma}}$ and $z_{2} = pf_{1} + (1-p)f_{0}$, but the difference $z_{2} - z_{1}$ is difficult to simplify. 

The second inequality \eqref{eqn:sqrt2} follows immediately from \eqref{eqn:sqrt1} by applying 
\[
\sqrt{pf_{1}} + \sqrt{(1-p)f_{0}} \leq \left(\sqrt{p} + \sqrt{1-p}\right) \sqrt{\max\{f_{0},f_{1}\}}
\]
to the denominator of the last term.
\end{IEEEproof}

Applying Lemma \ref{lem:sqrt} (specifically \eqref{eqn:sqrt2}) pointwise to the integrand in \eqref{eqn:Cp} 
results in  
\begin{align}
C_{p}^{1/2} &\leq \sqrt{p} \int_{-\infty}^{\infty} f_{1}(y) dy + \sqrt{1-p} \int_{-\infty}^{\infty} \sqrt{f_{0}(y) f_{1}(y)} dy\nonumber\\
&\qquad\quad - \frac{\sqrt{p(1-p)}}{\sqrt{p} + \sqrt{1-p}} \int_{-\infty}^{\infty} \sqrt{f_{1}(y)} \sqrt{\min\{ f_{0}(y), f_{1}(y) \}} dy\nonumber\\
&= \sqrt{p} +  C_{0}^{1/2} \sqrt{1-p} - \frac{\sqrt{p(1-p)}}{\sqrt{p} + \sqrt{1-p}} \left( \int_{\mcY'_{1}} \sqrt{f_{0}(y) f_{1}(y)} dy + \int_{\mcY'_{0}} f_{1}(y) dy \right)\nonumber\\
&= \frac{p}{\sqrt{p} + \sqrt{1-p}} + C_{0}^{1/2} \frac{1-p}{\sqrt{p} + \sqrt{1-p}} + \frac{\sqrt{p(1-p)}}{\sqrt{p} + \sqrt{1-p}} \left( \int_{\mcY'_{0}} \sqrt{f_{0}(y) f_{1}(y)} dy + \int_{\mcY'_{1}} f_{1}(y) dy \right)\nonumber\\
&= \frac{p}{\sqrt{p} + \sqrt{1-p}} + \frac{\sqrt{p(1-p)}}{\sqrt{p} + \sqrt{1-p}} \mbbP_{1}(\mcY'_{1}) + C_{0}^{1/2} \left( \frac{1-p}{\sqrt{p} + \sqrt{1-p}} + \frac{\sqrt{p(1-p)}}{\sqrt{p} + \sqrt{1-p}} \mbbP_{01}(\mcY'_{0}) \right).\label{eqn:BCpUB}
\end{align}
In the first equality above, we have used the definitions of $\mcY'_{0}$ and $\mcY'_{1}$ in the proposition statement.  
The second and third equalities follow from the complementarity of $\mcY'_{0}$ and $\mcY'_{1}$ and \eqref{eqn:C0Partial}.  
%
%
Combining \eqref{eqn:BCpUB} 
and the trivial bound $C_{p}^{1/2} \leq 1$ completes the proof. 
Since the regions $\mcY'_{0}$ and $\mcY'_{1}$ correspond to $\mcY_{0}$ and $\mcY_{1}$ in Proposition \ref{prop:Cp} with $p = 1/2$, the probabilities $\mbbP_{1}(\mcY'_{1})$ and $\mbbP_{01}(\mcY'_{0})$ are given by the same formulas as in \eqref{eqn:P1}, \eqref{eqn:P01}, \eqref{eqn:z1}, \eqref{eqn:z01} but with $\eta = \log\frac{1-p}{p} = 0$.
\end{IEEEproof}

\section{Comparison of upper bounds on the Bhattacharyya coefficient $C_{p}^{1/2}$}
\label{app:CpBCp}

\begin{figure}[t]
\centering
\includegraphics[width=0.4\textwidth]{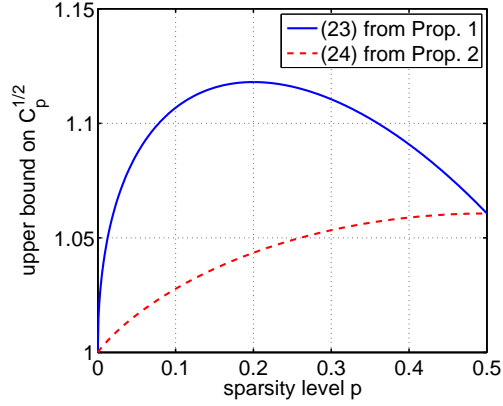}
\caption{Comparison of upper bounds \eqref{eqn:CpUBr0} and \eqref{eqn:BCpUBr0} on the Bhattacharyya coefficient $C_{p}^{1/2}$ at $r = 0$ as a function of $p$.}
\label{fig:BCpUBr0}
\end{figure}

In this appendix, we show that the bound in Proposition \ref{prop:BCp} on the Bhattacharyya coefficient $C_{p}^{1/2}$ is tighter than the first bound \eqref{eqn:CpUB1} in Proposition \ref{prop:Cp} with $\gamma = 1/2$ 
in the limit $r\lambda \to 0$.  We restrict attention to the case $p < 1/2$ so that $\eta = \log\frac{1-p}{p} > 0$.  Then it can be seen from \eqref{eqn:z1}--\eqref{eqn:z0} that $z_{1}^{\pm}, z_{0}^{\pm} \to -\infty$ as $r\lambda \to 0$ while $z_{01}^{\pm} \to \pm \infty$.  Hence \eqref{eqn:P1}--\eqref{eqn:P0} imply that $\lim_{r\lambda\to 0} \mbbP_{1}(\mcY_{1}) = \lim_{r\lambda\to 0} \mbbP_{0}(\mcY_{1}) = 0$ and $\lim_{r\lambda\to 0} \mbbP_{01}(\mcY_{0}) = 1$.  Combined with the limit $C_{0}^{1/2} \to 1$ from \eqref{eqn:C0}, the non-trivial bound in \eqref{eqn:CpUB1} in Proposition \ref{prop:Cp} evaluates to 
\begin{equation}\label{eqn:CpUBr0}
\frac{1}{2} \sqrt{p} + \sqrt{1-p} 
\end{equation}
for $\gamma = 1/2$.  For the bound in Proposition \ref{prop:BCp}, we have instead $\eta = 0$ in \eqref{eqn:z1}, \eqref{eqn:z01}.  While it is still the case that $z_{1}^{-}, z_{01}^{-} \to -\infty$ as $r\lambda \to 0$, now $z_{1}^{+}, z_{01}^{+} \to 0$ because the quantities in parentheses in \eqref{eqn:z1}, \eqref{eqn:z01} decrease to zero as $O(r\lambda)$ while the prefactor is $O(1/\sqrt{r\lambda})$.  It follows that $\lim_{r\lambda\to 0} \mbbP_{1}(\mcY'_{1}) = \lim_{r\lambda\to 0} \mbbP_{01}(\mcY'_{0}) = 1/2$ and the bound in Proposition \ref{prop:BCp} becomes 
\begin{equation}\label{eqn:BCpUBr0}
\frac{1 + \sqrt{p(1-p)}}{\sqrt{p} + \sqrt{1-p}}.
\end{equation}
Fig.~\ref{fig:BCpUBr0} compares \eqref{eqn:CpUBr0} and \eqref{eqn:BCpUBr0} and shows that the latter is smaller for all $p \in (0, 1/2)$.  While it is true that both \eqref{eqn:CpUBr0} and \eqref{eqn:BCpUBr0} exceed the trivial upper bound of $1$, the fact that \eqref{eqn:BCpUBr0} is closer to $1$ implies that the trivial bound is improved upon sooner as $r\lambda$ increases from zero and the Bhattacharyya coefficient $C_{0}^{1/2}$ decays from $1$. 

\section{Proof of Theorem \ref{thm:pFixedHighSNR}}
\label{app:pfpFixedHighSNR}

The theorem stems from an asymptotic expansion of the bounds in Proposition \ref{prop:Cp} on the Chernoff coefficient $C_{p}^{\gamma}$  
in the limit $r\lambda \to \infty$.  

\begin{lemma}\label{lem:CpHighSNR}
As $r\lambda \to \infty$, the Chernoff coefficient $C_{p}^{\gamma}$ is governed by the following expansion:
\[
C_{p}^{\gamma} \leq p^{1-\gamma} \left(1 + \left(\frac{1-p}{p}\right)^{1-\gamma} \frac{e^{-\gamma s/2}}{\sqrt{1-\gamma}} \frac{1}{(r\lambda)^{\gamma/2}} - O \left( \sqrt{\frac{\log(r\lambda)}{r\lambda}} \right) \right).
\]
\end{lemma}
\begin{IEEEproof}
We begin with the first bound \eqref{eqn:CpUB1} in Proposition \ref{prop:Cp}.  There are four quantities to consider: $C_{0}^{\gamma}$, $\mbbP_{1}(\mcY_{1})$, $\mbbP_{01}(\mcY_{0})$, and $\mbbP_{0}(\mcY_{1})$, and we expand each to lowest order in $1/r\lambda$.  From \eqref{eqn:C0} we have 
\begin{align}
C_{0}^{\gamma} &= \frac{(r\lambda)^{-\gamma/2}}{\sqrt{1-\gamma}} \sqrt{\frac{\left(1 + \frac{1}{r\lambda}\right)^{1-\gamma}}{1 + \frac{1}{(1-\gamma)r\lambda}}} \exp\left( -\frac{\gamma s}{2} + \frac{\gamma s}{2(1 + (1-\gamma)r\lambda)} \right)\nonumber\\
&= \frac{e^{-\gamma s/2}}{\sqrt{1-\gamma}} \frac{1}{(r\lambda)^{\gamma/2}} \left(1 + O\left( \frac{1}{r\lambda}\right)\right).\label{eqn:C0HighSNR}
\end{align}
Next we express $z_{1}^{\pm}$ in \eqref{eqn:z1} as 
\begin{align}
z_{1}^{\pm} &= \pm\sqrt{s} \sqrt{1 + \frac{1}{r\lambda}} - \sqrt{\frac{s + 2\eta + \log(1+r\lambda)}{r\lambda}}\nonumber\\
&= \pm\sqrt{s} - \sqrt{\frac{s + 2\eta + \log(r\lambda)}{r\lambda}} + O\left( \frac{1}{r\lambda} \right).\label{eqn:z1HighSNR}
\end{align}
Substituting \eqref{eqn:z1HighSNR} into \eqref{eqn:P1} and expanding to first order, 
\begin{align}
\mbbP_{1}(\mcY_{1}) &= \Phi(\sqrt{s}) - \phi(\sqrt{s}) \sqrt{\frac{s + 2\eta + \log(r\lambda)}{r\lambda}}\nonumber\\
&\qquad\quad + \Phi(-\sqrt{s}) - \phi(-\sqrt{s}) \sqrt{\frac{s + 2\eta + \log(r\lambda)}{r\lambda}} + O\left( \frac{\log(r\lambda)}{r\lambda} \right)\nonumber\\
&= 1 - O\left( \sqrt{\frac{\log(r\lambda)}{r\lambda}} \right),\label{eqn:P1HighSNR}
\end{align}
where $\phi$ is the standard Gaussian PDF.  For $z_{01}^{\pm}$ in \eqref{eqn:z01} we have 
\begin{align}
z_{01}^{\pm} &= \sqrt{s(1-\gamma)} \sqrt{1 + \frac{1}{(1-\gamma)r\lambda}} \left(-\frac{\sqrt{1 + r\lambda}}{1 + (1-\gamma) r\lambda} \pm \sqrt{1 + \frac{1}{s} (2\eta + \log(1+r\lambda))} \right)\nonumber\\
&= \pm\sqrt{(1-\gamma) (s + 2\eta + \log(r\lambda))} - O\left( \frac{1}{\sqrt{r\lambda}} \right),\label{eqn:z01HighSNR}
\end{align}
which approaches $\pm\infty$ as $r\lambda \to \infty$.  Using the Gaussian tail bounds $\Phi(z) > 1 - \phi(z)/z$ for $z > 0$ and $\Phi(z) < -\phi(z)/z$ for $z < 0$, we infer from \eqref{eqn:P01} and \eqref{eqn:z01HighSNR} that 
\begin{align}
\mbbP_{01}(\mcY_{0}) &= \Phi\left( \sqrt{(1-\gamma) (s + 2\eta + \log(r\lambda))} \right) - \Phi\left( -\sqrt{(1-\gamma) (s + 2\eta + \log(r\lambda))} \right)\nonumber\\
&\qquad\quad + \phi\left( \sqrt{(1-\gamma) (s + 2\eta + \log(r\lambda))} \right) O\left( \frac{1}{\sqrt{r\lambda}} \right)\nonumber\\
&> 1 + \phi\left( \sqrt{(1-\gamma) (s + 2\eta + \log(r\lambda))} \right) \left( -\frac{2}{\sqrt{(1-\gamma) (s + 2\eta + \log(r\lambda))}} + O\left( \frac{1}{\sqrt{r\lambda}} \right) \right)\nonumber\\
&= 1 + \frac{1}{\sqrt{2\pi}} \exp\left( -\frac{1-\gamma}{2} (s + 2\eta + \log(r\lambda)) \right) \left( -\frac{2}{\sqrt{(1-\gamma) (s + 2\eta + \log(r\lambda))}} + O\left( \frac{1}{\sqrt{r\lambda}} \right) \right)\nonumber\\
&= 1 - O\left( \frac{1}{(r\lambda)^{(1-\gamma)/2} \sqrt{\log(r\lambda)}} \right).\label{eqn:P01HighSNR}
\end{align}
%
Lastly we consider $z_{0}^{\pm}$ in \eqref{eqn:z0}:
\begin{align*}
z_{0}^{\pm} &= -\sqrt{\left(1 + \frac{1}{r\lambda}\right)(s + 2\eta + \log(1+r\lambda))} \pm \sqrt{\frac{s}{r\lambda}}\\
&= -\sqrt{s + 2\eta + \log(r\lambda)} \pm \sqrt{\frac{s}{r\lambda}} + O\left( \frac{1}{r\lambda} \right),
\end{align*}
which leads to 
\begin{align}
\mbbP_{0}(\mcY_{1}) &= 2\Phi\left( -\sqrt{s + 2\eta + \log(r\lambda)} \right) + \phi\left( -\sqrt{s + 2\eta + \log(r\lambda)} \right) \left(\sqrt{\frac{s}{r\lambda}} - \sqrt{\frac{s}{r\lambda}} + O\left( \frac{1}{r\lambda} \right) \right)\nonumber\\
&< \frac{1}{\sqrt{2\pi}} \exp\left( -\frac{1}{2} (s + 2\eta + \log(r\lambda)) \right) \left( \frac{2}{\sqrt{s + 2\eta + \log(r\lambda)}} + O\left( \frac{1}{r\lambda} \right) \right)\nonumber\\
&= O\left( \frac{1}{\sqrt{r\lambda \log(r\lambda)}} \right),\label{eqn:P0HighSNR}
\end{align}
using the tail bound $\Phi(z) < -\phi(z)/z$ again in the second line.
Combining \eqref{eqn:C0HighSNR}, \eqref{eqn:P1HighSNR}, \eqref{eqn:P01HighSNR}, and \eqref{eqn:P0HighSNR} with \eqref{eqn:CpUB1}, we deduce that for large $r\lambda$, 
\begin{align*}
C_{p}^{\gamma} &\leq p^{1-\gamma} \left(1 - O\left( \sqrt{\frac{\log(r\lambda)}{r\lambda}} \right) \right)\\
&\qquad\quad + (1-p)^{1-\gamma} \frac{e^{-\gamma s/2}}{\sqrt{1-\gamma}} \frac{1}{(r\lambda)^{\gamma/2}} \left(1 - O\left( \frac{1}{(r\lambda)^{(1-\gamma)/2} \sqrt{\log(r\lambda)}} \right) \right)\\
&\qquad\quad + O\left( \frac{1}{\sqrt{r\lambda \log(r\lambda)}} \right)\\
&= p^{1-\gamma} + (1-p)^{1-\gamma} \frac{e^{-\gamma s/2}}{\sqrt{1-\gamma}} \frac{1}{(r\lambda)^{\gamma/2}} - O\left( \sqrt{\frac{\log(r\lambda)}{r\lambda}} \right). 
\end{align*}
\end{IEEEproof}

From Lemma \ref{lem:CpHighSNR} we immediately obtain 
%
\begin{align}
\left(C_{p}^{\gamma}\right)^{-\frac{1}{1-\gamma}} - 1 &\geq \frac{1}{p} \left(1 + \left(\frac{1-p}{p}\right)^{1-\gamma} \frac{e^{-\gamma s/2}}{\sqrt{1-\gamma}} \frac{1}{(r\lambda)^{\gamma/2}} - O \left( \sqrt{\frac{\log(r\lambda)}{r\lambda}} \right) \right)^{-\frac{1}{1-\gamma}} - 1\nonumber\\
&= \frac{1}{p} \left(1 - \left(\frac{1-p}{p}\right)^{1-\gamma} \frac{e^{-\gamma s/2}}{(1-\gamma)^{3/2}} \frac{1}{(r\lambda)^{\gamma/2}} + O \left( \sqrt{\frac{\log(r\lambda)}{r\lambda}} \right) \right) - 1\nonumber\\
&= \frac{1-p}{p} \left(1 - \frac{1}{p^{1-\gamma} (1-p)^{\gamma}} \frac{e^{-\gamma s/2}}{(1-\gamma)^{3/2}} \frac{1}{(r\lambda)^{\gamma/2}} + O \left( \sqrt{\frac{\log(r\lambda)}{r\lambda}} \right) \right).\label{eqn:CpHighSNR}
\end{align}
%
Substituting \eqref{eqn:CpHighSNR} into the bound on $G$ in Theorem \ref{thm:pFixed} gives 
\begin{align}
\lim_{N\to\infty} G &\geq \left[ 1 + \frac{r}{r+1} \frac{1-p}{p} \max_{\lambda\in[0,1]} \left(1 - \frac{1}{p^{1-\gamma} (1-p)^{\gamma}} \frac{e^{-\gamma s/2}}{(1-\gamma)^{3/2}} \frac{1}{(r\lambda)^{\gamma/2}} + O \left( \sqrt{\frac{\log(r\lambda)}{r\lambda}} \right) \right) (1-\lambda) \right]^{q/2}\nonumber\\
&= \left[ 1 + \frac{r}{r+1} \frac{1-p}{p} \max_{\lambda\in[0,1]} 1 - \lambda - \frac{1}{p^{1-\gamma} (1-p)^{\gamma}} \frac{e^{-\gamma s/2}}{(1-\gamma)^{3/2}} \frac{1}{(r\lambda)^{\gamma/2}} + O \left( \sqrt{\frac{\log(r\lambda)}{r\lambda}} + \frac{\lambda}{(r\lambda)^{\gamma/2}} \right) \right]^{q/2}.\label{eqn:GHighSNR}
\end{align}
The neglect of the cross term $O(\lambda / (r\lambda)^{\gamma/2})$ will be justified shortly.  
Noting that the leading terms in $\lambda$ specify a concave function, 
the maximization over $\lambda$ in \eqref{eqn:GHighSNR} can now be carried out analytically by setting the derivative with respect to $\lambda$ to zero and solving for the optimal $\lambda^{\ast}$, resulting in
\[
\lambda^{\ast} = \frac{(\gamma/2)^{\frac{2}{2+\gamma}}}{(1-\gamma)^{\frac{3}{2+\gamma}}} \frac{1}{p^{\frac{2(1-\gamma)}{2+\gamma}} (1-p)^{\frac{2\gamma}{2+\gamma}}} e^{-\frac{\gamma s}{2+\gamma}} r^{-\frac{\gamma}{2+\gamma}}.
\]
Substituting $\gamma = 2/(q+2)$ and simplifying yields \eqref{eqn:lambda*HighSNR}.  Since $\lambda^{\ast} \propto r^{-\frac{\gamma}{2+\gamma}}$ and $(r\lambda^{\ast})^{-\gamma/2} \propto r^{-\frac{\gamma}{2+\gamma}}$, the neglected cross term $\lambda / (r\lambda)^{\gamma/2} \propto r^{-\frac{2\gamma}{2+\gamma}}$ is of higher order as claimed.  We now substitute $\lambda = \lambda^{\ast}$ into \eqref{eqn:GHighSNR}, noting that the prefactor $r/(r+1) = 1 - O(1/r)$: 
\begin{align*}
\lim_{N\to\infty} G &\geq \left[ 1 + \frac{1-p}{p} \left( 1 - \lambda^{\ast} - \frac{2}{\gamma} \lambda^{\ast} + O\left( r^{-\frac{1}{2+\gamma}} \sqrt{\log r} + r^{-\frac{2\gamma}{2+\gamma}} \right) \right) \right]^{q/2}\\
&= \left[ \frac{1}{p} \left( 1 - (1-p) \frac{2+\gamma}{\gamma} \lambda^{\ast} + O\left( r^{-\frac{\min\{1, 2\gamma\}}{2+\gamma}} \sqrt{\log r} \right) \right) \right]^{q/2}\\
&= \left( \frac{1}{p} \right)^{q/2} \left[ 1 - (q+3)(1-p) \lambda^{\ast} + O\left( r^{-\frac{1+\min\{q/2,1\}}{q+3}} \sqrt{\log r} \right) \right]^{q/2}\\
&= \left( \frac{1}{p} \right)^{q/2} \left[ 1 - \frac{q(q+3)}{2} (1-p) \lambda^{\ast} + O\left( r^{-\frac{1+\min\{q/2,1\}}{q+3}} \sqrt{\log r} \right) \right],
\end{align*}
again using $\gamma = 2/(q+2)$ in the second equality.  Combining with \eqref{eqn:lambda*HighSNR} completes the proof.
\qquad\endproof

\section{Proof of Theorem \ref{thm:pVanishing}}
\label{app:pfpVanishing}

Similar to Theorem \ref{thm:pFixedHighSNR}, the proof relies on Taylor expansions.  As $r \to 0$, we may expand $\left(C_{0}^{\gamma}\right)^{-\frac{1}{1-\gamma}}$ to first order in $r\lambda$ using \eqref{eqn:C0}:
\begin{align*}
\left(C_{0}^{\gamma}\right)^{-\frac{1}{1-\gamma}} &= \frac{(1 + (1-\gamma)r\lambda)^{\frac{1}{2(1-\gamma)}}}{\sqrt{1 + r\lambda}} \exp\left( \frac{\gamma sr\lambda}{2(1 +(1-\gamma) r\lambda)} \right)\\
&= \left(1 + \frac{r\lambda}{2} + O\left((r\lambda)^{2}\right) \right) \left(1 - \frac{r\lambda}{2} + O\left((r\lambda)^{2}\right) \right) \left(1 + \frac{\gamma sr\lambda}{2} + O\left((r\lambda)^{2}\right) \right)\\
&= 1 + \frac{\gamma sr\lambda}{2} + O\left((r\lambda)^{2}\right).
\end{align*}
Substituting into \eqref{eqn:G0} gives 
\[
\lim_{\substack{Np\to\infty\\p\to 0}} G \geq \left[ 1 + \frac{r}{r+1} \max_{\lambda\in[0,1]} \left( \frac{\gamma sr\lambda}{2} + O\left((r\lambda)^{2}\right) \right) (1-\lambda) \right]^{q/2}.
\]
The right-hand side is of the form $1 + c \lambda (1-\lambda)$ to lowest order in $r$ and is thus maximized at $\lambda^{\ast} = 1/2$.  Consequently we obtain 
\begin{align*}
\lim_{\substack{Np\to\infty\\p\to 0}} G &\geq \left[ 1 + \frac{r}{r+1} \left( \frac{\gamma s r}{8} + O\left((r\lambda)^{2}\right) \right) \right]^{q/2}\\
&= \left[ 1 + \frac{\gamma s r^{2}}{8} + O\left((r\lambda)^{3}\right) \right]^{q/2}\\
&= 1 + \frac{(1-\gamma) s r^{2}}{8} + O\left((r\lambda)^{3}\right),
\end{align*}
using $r / (r+1) = r - O(r^{2})$ in the first equality and $q\gamma/2 = 1-\gamma$ in the second. 

As $r \to \infty$, 
\begin{align*}
\left(C_{0}^{\gamma}\right)^{-\frac{1}{1-\gamma}} &= \frac{\left((1-\gamma) r\lambda\right)^{\frac{1}{2(1-\gamma)}} \left(1 + \frac{1}{(1-\gamma)r\lambda}\right)^{\frac{1}{2(1-\gamma)}}}{(r\lambda)^{1/2} \sqrt{1 + \frac{1}{r\lambda}}} \exp\left( \frac{\gamma s}{2(1-\gamma)} \frac{1}{1 + \frac{1}{(1-\gamma) r\lambda}} \right)\\
&= (1-\gamma)^{\frac{1}{2(1-\gamma)}} \exp\left( \frac{\gamma s}{2(1-\gamma)} \right) (r\lambda)^{\frac{\gamma}{2(1-\gamma)}} \left(1 + O\left( \frac{1}{r\lambda} \right) \right)\\
&= \left( \frac{q}{q+2} \right)^{\frac{q+2}{2q}} e^{s/q} (r\lambda)^{1/q} \left(1 + O\left( \frac{1}{r\lambda} \right) \right)
\end{align*}
using $\gamma = 2 / (q + 2)$. 
Substituting into \eqref{eqn:G0} yields 
\[
\lim_{\substack{Np\to\infty\\p\to 0}} G \geq \left[ 1 + \frac{r}{r+1} \max_{\lambda\in[0,1]} \left( \left( \frac{q}{q+2} \right)^{\frac{q+2}{2q}} e^{s/q} (r\lambda)^{1/q} + O\left((r\lambda)^{\max\{1/q-1,0\}}\right) \right) (1-\lambda) \right]^{q/2}.
\]
The leading term in $r$ is proportional to $\lambda^{1/q} (1-\lambda)$, which is maximized at $\lambda^{\ast} = 1/(q+1)$ as can be verified by differentiating with respect to $\lambda$.  Noting that $r/(r+1) = 1 - O(1/r)$, this leads to 
\begin{align*}
\lim_{\substack{Np\to\infty\\p\to 0}} G &\geq \left[ 1 + \left( 1 - O\left(\frac{1}{r}\right) \right) \left( \left( \frac{q}{q+2} \right)^{\frac{q+2}{2q}} \left(\frac{1}{q+1}\right)^{1/q} \frac{q}{q+1} e^{s/q} r^{1/q} + O\left(r^{\max\{1/q-1,0\}}\right) \right) \right]^{q/2}\\
&= \left[ \frac{q^{\frac{3q+2}{2q}}}{(q+2)^{\frac{q+2}{2q}} (q+1)^{\frac{q+1}{q}}} e^{s/q} r^{1/q} + O\left(r^{\max\{1/q-1,0\}}\right) \right]^{q/2}\\
&= \frac{q^{\frac{3q+2}{4}}}{(q+2)^{\frac{q+2}{4}} (q+1)^{\frac{q+1}{2}}} e^{s/2} \sqrt{r} + O\left(r^{\max\left\{\frac{1}{2}-\frac{1}{q},-\frac{1}{2}\right\}}\right).
\end{align*}
%
\qquad\endproof



\ifCLASSOPTIONcaptionsoff
  \newpage
\fi



\bibliographystyle{IEEEtran}
\bibliography{IEEEabrv,adapSens,sparseLinInv,optimization}

\begin{thebibliography}{10}
\providecommand{\url}[1]{#1}
\csname url@samestyle\endcsname
\providecommand{\newblock}{\relax}
\providecommand{\bibinfo}[2]{#2}
\providecommand{\BIBentrySTDinterwordspacing}{\spaceskip=0pt\relax}
\providecommand{\BIBentryALTinterwordstretchfactor}{4}
\providecommand{\BIBentryALTinterwordspacing}{\spaceskip=\fontdimen2\font plus
\BIBentryALTinterwordstretchfactor\fontdimen3\font minus
  \fontdimen4\font\relax}
\providecommand{\BIBforeignlanguage}[2]{{%
\expandafter\ifx\csname l@#1\endcsname\relax
\typeout{** WARNING: IEEEtran.bst: No hyphenation pattern has been}%
\typeout{** loaded for the language `#1'. Using the pattern for}%
\typeout{** the default language instead.}%
\else
\language=\csname l@#1\endcsname
\fi
#2}}
\providecommand{\BIBdecl}{\relax}
\BIBdecl

\bibitem{adapEstJSTSP2013}
D.~Wei and A.~O. Hero, ``Multistage adaptive estimation of sparse signals,''
  \emph{IEEE J. Sel. Topics Signal Process.}, vol.~7, no.~5, pp. 783--796, Oct.
  2013.

\bibitem{bashan2008}
E.~Bashan, R.~Raich, and A.~O. Hero, ``Optimal two-stage search for sparse
  targets using convex criteria,'' \emph{{IEEE} Trans. Signal Process.},
  vol.~56, pp. 5389--5402, Nov. 2008.

\bibitem{hitchings2010}
D.~Hitchings and D.~A. Casta{\~{n}}\'{o}n, ``Adaptive sensing for search with
  continuous actions and observations,'' in \emph{Proc. IEEE Conf. Decision and
  Control (CDC)}, Dec. 2010, pp. 7443--7448.

\bibitem{bashan2011}
E.~Bashan, G.~Newstadt, and A.~O. Hero, ``Two-stage multiscale search for
  sparse targets,'' \emph{{IEEE} Trans. Signal Process.}, vol.~59, pp.
  2331--2341, May 2011.

\bibitem{haupt2011}
J.~Haupt, R.~M. Castro, and R.~Nowak, ``Distilled sensing: Adaptive sampling
  for sparse detection and estimation,'' \emph{{IEEE} Trans. Inf. Theory},
  vol.~57, pp. 6222--6235, Sep. 2011.

\bibitem{malloy2011b}
M.~Malloy and R.~Nowak, ``On the limits of sequential testing in high
  dimensions,'' in \emph{Conf. Rec. Asilomar Conf. Signals Syst. Comput.}, Nov.
  2011, pp. 1245--1249.

\bibitem{tajer2012}
A.~Tajer, R.~M. Castro, and X.~Wang, ``Adaptive sensing of congested spectrum
  bands,'' \emph{{IEEE} Trans. Inf. Theory}, vol.~58, no.~9, pp. 6110--6125,
  Sep. 2012.

\bibitem{iwen2012}
M.~A. Iwen and A.~H. Tewfik, ``Adaptive strategies for target detection and
  localization in noisy environments,'' \emph{{IEEE} Trans. Signal Process.},
  vol.~60, no.~5, pp. 2344--2353, May 2012.

\bibitem{haupt2012}
J.~Haupt, R.~Baraniuk, R.~Castro, and R.~Nowak, ``Sequentially designed
  compressed sensing,'' in \emph{Proc. IEEE Statist. Signal Process. Workshop
  (SSP)}, Aug. 2012, pp. 1--4.

\bibitem{malloy2012}
M.~L. Malloy and R.~D. Nowak, ``Near-optimal adaptive compressed sensing,'' in
  \emph{Conf. Rec. Asilomar Conf. Signals Syst. Comput.}, Nov. 2012, pp.
  1935--1939.

\bibitem{ji2008}
S.~Ji, Y.~Xue, and L.~Carin, ``Bayesian compressive sensing,'' \emph{{IEEE}
  Trans. Signal Process.}, vol.~56, pp. 2346--2356, Jun. 2008.

\bibitem{castro2008}
R.~M. Castro, J.~Haupt, R.~Nowak, and G.~M. Raz, ``Finding needles in noisy
  haystacks,'' in \emph{Proc. IEEE Int. Conf. Acoust. Speech Signal Process.
  (ICASSP)}, Apr. 2008, pp. 5133--5136.

\bibitem{bertsekas2005}
D.~P. Bertsekas, \emph{Dynamic Programming and Optimal Control}, 3rd~ed.\hskip
  1em plus 0.5em minus 0.4em\relax Nashua, NH: Athena Scientific, 2005, vol.~1.

\bibitem{jenkins2010}
K.~L. Jenkins and D.~A. Casta{\~{n}}\'{o}n, ``Adaptive sensor management for
  feature-based classification,'' in \emph{Proc. IEEE Conf. Decision and
  Control (CDC)}, Dec. 2010, pp. 522--527.

\bibitem{jenkins2011}
------, ``Information-based adaptive sensor management for sensor networks,''
  in \emph{Proc. American Control Conf. (ACC)}, Jun. 2011, pp. 4934--4940.

\bibitem{firouzi2013}
H.~Firouzi, B.~Rajaratnam, and A.~O. Hero, ``Predictive correlation screening:
  Application to two-stage predictor design in high dimension,'' Apr. 2013,
  arXiv:1301.2378.

\bibitem{auer2002}
P.~Auer, N.~Cesa-Bianchi, and P.~Fischer, ``Finite-time analysis of the
  multiarmed bandit problem,'' \emph{Machine Learning}, vol.~47, no. 2--3, pp.
  235--256, May 2002.

\bibitem{audibert2007}
J.-Y. Audibert, R.~Munos, and C.~Szepesv\'{a}ri, ``Tuning bandit algorithms in
  stochastic environments,'' in \emph{Algorithmic Learning Theory}.\hskip 1em
  plus 0.5em minus 0.4em\relax Springer, 2007, pp. 150--165.

\bibitem{adapEstICASSP2013}
D.~Wei and A.~O. Hero, ``Adaptive spectrum sensing and estimation,'' in
  \emph{Proc. IEEE Int. Conf. Acoust. Speech Signal Process. (ICASSP)},
  Vancouver, Canada, May 2013, pp. 5720--5724.

\bibitem{wu2011}
Y.~Wu and S.~Verd\'{u}, ``{MMSE} dimension,'' \emph{{IEEE} Trans. Inf. Theory},
  vol.~57, no.~8, pp. 4857--4879, Aug. 2011.

\bibitem{ariascastro2013}
E.~Arias-Castro, E.~J. Candes, and M.~A. Davenport, ``On the fundamental limits
  of adaptive sensing,'' \emph{{IEEE} Trans. Inf. Theory}, vol.~59, no.~1, pp.
  472--481, Jan. 2013.

\bibitem{castro2013}
R.~M. Castro, ``Adaptive sensing performance lower bounds for sparse signal
  estimation and testing,'' Mar. 2013, arXiv:1206.0648.

\bibitem{vandervaart2000}
A.~W. van~der Vaart, \emph{Asymptotic Statistics}, ser. Cambridge Series in
  Statistical and Probabilistic Mathematics.\hskip 1em plus 0.5em minus
  0.4em\relax Cambridge University Press, 2000.

\end{thebibliography}
\end{document}